\RequirePackage{fix-cm}
\documentclass[epj]{svjour}
\smartqed  % flush right qed marks, e.g. at end of proof
\RequirePackage{graphicx}
\usepackage{graphicx} % Include figure files
\usepackage{amsmath}
\usepackage{amssymb}
\usepackage{dcolumn} % Align table columns on decimal point
\usepackage{multicol} % multicol for Table II footnotes
\usepackage{bm} % bold math
\usepackage{xcolor} % color text block package (SSaha added for modification tag 17/5/20)
\def\beq{\begin{equation}}
\def\eeq{\end{equation}}
\def\bea{\begin{eqnarray}}
\def\eea{\end{eqnarray}}

\usepackage{threeparttable} % for table with footnotes.

\begin{document}

\title{High-spin states of $^{204}$At: isomeric states and shears band structure%\thanksref{t1}
}


\author{D. Kanjilal\inst{1}
        \and
        S. K. Dey\inst{2}\,\inst{3}
        \and
        S. S. Bhattacharjee\inst{8}
        \and
        A. Bisoi\inst{6}
        \and
        M. Das\inst{2}
        \and
        C.~C.~Dey\inst{2}
        \and
        S. Nag\inst{7}
        \and \\
        R. Palit\inst{4}
	    \and
        S. Ray\inst{5}
        \and
        S. Saha\inst{4}
        \and
        J. Sethi\inst{4}
        \and
        S.~Saha\inst{2}
\thanks{\emph{e-mail: satyajit.saha@saha.ac.in}}
}

%\thankstext{e1}{e-mail: satyajit.saha@saha.ac.in}


\institute{Department of Physics, Raiganj Surendranath Mahavidyalaya, Raiganj, West Bengal 733134, India
           \and
           Saha Institute of Nuclear Physics, A CI of Homi Bhabha National Institute, I/AF Bidhan Nagar, Kolkata 700064, India
           \and
           \emph{Present Affiliation: KEK, Japan}
           \and
           Tata Institute of Fundamental Research, Mumbai 400005, India
           \and
           Mody University of Science and Technology, Sikar, Rajasthan 332311, India
           \and
           Indian Institute of Engineering Science and Technology, Shibpur, Howrah 711103, India
           \and
           Indian Institute of Technology (Banaras Hindu University), Varanasi 221005, India
           \and
           TRIUMF, 4004 Wesbrook Mall, Vancouver, British Columbia V6T 2A3, Canada
}

\date{Received: date / Accepted: date}

\abstract{
High-spin states of neutron deficient Trans-Lead nucleus $^{204}$At were populated up to $\sim 8\,{\rm MeV}$ excitation through the $^{12}$C + $^{197}$Au fusion evaporation reaction. Decay of the associated levels through prompt and delayed $\gamma$-ray emissions were studied to evaluate the underlying nuclear structure. The level scheme, which was partly known, was extended further. An isomeric $16^+$ level with observed lifetime $\tau=52 \pm 5\, {\rm ns}$, was established from our measurements. Attempts were made to interpret the excited states based on multi quasiparticle and hole structures involving $2f_{5/2}$, $1h_{9/2}$, {\rm and} $1i_{13/2}$ shell model states, along with moderate core excitation. Magnetic dipole band structure over the spin parity range:~$16^+ - 23^+$ was confirmed and evaluated in more detail, including the missing cross-over $E2$ transitions. Band-crossing along the shears band was observed and compared with the evidence of similar phenomena in the neighbouring neutron deficient $^{202}$Bi, $^{205}$Rn isotones and the $^{203}$At isotope. Based on comparison of the measured $B(M1)/B(E2)$ values for transitions along the band with the semiclassical model based estimates, the shears band of $^{204}$At was established along with the level scheme.
} 

%\keywords{NUCLEAR STRUCTURE $^{204}$At \and excitation to high spin levels by $^{197}$Au($^{12}$C,$5n$) \and $E_\gamma$, $I_\gamma$,  $\gamma \gamma$-coin, 
%$\gamma \gamma (\theta)$-DCO and $\gamma \gamma$ (linear polarization)-PDCO \and deduced levels, $J^\pi$, multipolarity \and isomer decay \and quasiparticle configurations \and magnetic dipole rotational band \and deduced $B(M1)/B(E2)$ \and SPAC model}


\maketitle
%
% Computer program descriptions must contain the following
% PROGRAM SUMMARY AND SPECIFICATIONS.
%\noindent
%{\bf Program Summary and Specifications}\\
%Delete as appropriate.
%\begin{small}
%\noindent
%{Program title:}\\
%{Licensing provisions:}\\
%Please choose one among the following:  CC0 1.0/CC By 4.0/MIT/Apache-2.0/BSD 3-clause/BSD 2-clause/GPLv3/GPLv2/LGPL/CC BY NC 3.0/MPL-2.0    
%{Programming language:}\\
%{Repository and DOI:}\\
%Please indicate where the program and its additional files have been deposited (e.g. Github,…) , together with the DOI obtained. 
%{Description of problem:}\\
%Describe the nature of the problem here in approx. 50-250 words. 
%{Method of solution:}\\
%Describe the method solution here in approx. 50-250 words. 
%{Additional comments:}\\
%Provide any additional comments (e.g. previous versions, new version information and summary, unusual features, limitations,…) here in approx. 50-250 words.
%\end{small}
%
\section{Introduction}
\label{intro}
The nuclear structure of the neutron deficient nuclei very near the doubly magic $^{208}$Pb nucleus has been one of the major areas of experimental investigation for many reasons. Firstly, many of these nuclei were studied to look for applicability of shell model with moderate core excitation to explain the high-spin states\cite{mar,Pal,Fan2}. These studies were possible due to the availability of cooled High Purity Germanium (HPGe) based $\gamma$-ray detectors with unprecedented energy resolution to pin point the basic structural subtleties. Secondly, the yield of neutron deficient trans-Lead nuclei populated to high-spin states by fusion evaporation pathway is very low due to depletion of the compound nuclei by the competing fission channels. However, large array of HPGe detectors and the Clover detectors made available over the last three decades, along with versatile techniques of channel selection, made it possible to probe these nuclei to very high spin and excitation energy. Thirdly, these nuclei with a few valence protons and neutron holes, which belong to relatively high-$j$ orbitals ($f_{5/2}, h_{9/2}$, and $i_{13/2}$), are expected to manifest various co-operative phenomena as the collectivity sets in for the high-spin states. 

With moderate core excitations as the basis, these nuclei $(Z \sim 82, N \sim 120)$ first evolve from a spherical to weak oblate shape. The valence particles and holes tend to align along the rotational symmetry axis giving rise to evolution of collective phenomena, the simplest manifestation being the observation of a series of magnetic dipole transitions between the high-spin states which appear to be regularly or semi-regularly placed following some order pattern. The magnetic dipole band, interpreted physically by the shears mechanism, has been explained by the tilted axis cranking (TAC) model\cite{frau,frau1}. Shape co-existence and shape transition due to transformation from weak oblate shape to prolate shape occurs as the neutron number decreases further\cite{203at}. This was first experimentally observed in this mass region in $^{186}$Pb\cite{andr}. These were also observed in the neutron deficient Polonium isotopes produced in the high-spin states by Coulomb excitation of post accelerated beams of Polonium isotopes at the REX-ISOLDE facility\cite{kest}. Theoretical attempts were also made using Relativistic Mean Field (RMF) and other many body techniques to explain the results. Exclusive RMF calculations were also done to predict the shape evolution in the neutron deficient Astatine(At) isotopes\cite{lian}. Multiple minima on the potential energy surfaces (PES) predict shape coexistence for At isotopes around $A \sim 200$. The shape stabilizes from weak oblate around $A \sim 204$ towards spherical shape at larger $A$ values.

Magnetic dipole bands were first observed in neutron deficient isotopes of Lead\cite{clar1}. Following these studies, a number of magnetic dipole bands were discovered in nuclei around Lead with onset of weak oblate deformation\cite{clar2,Hub1}. Only a few magnetic dipole bands were found in nuclei for $Z=84$ and above. These are: $^{205}$Rn\cite{nova}, $^{206}$Fr and $^{204}$At\cite{hart} and recently in $^{201}$At\cite{201at} and $^{203}$At\cite{203at}. In almost all the cases, these shears bands are created based on the coupling of a few high-spin protons involving $h_{9/2}$ and $i_{13/2}$ orbitals and the neutron holes in the $i_{13/2} $ sub shell. A previous study of the high-spin states of the $^{204}$At\cite{hart} nucleus using Gammasphere revealed the dipole band, however, the corresponding spin parities were not assigned due to limited statistics on the neighbouring transitions, which are either feeding the states belonging to the dipole band or fed by these states. Furthermore, cross over $E2$ type intra-band transitions could not be observed or studied. Qualitative or quantitative estimates based on the TAC or other empirical models could not be done to understand or compare the band structure with theoretical predictions.

In this paper, we have attempted to establish the level scheme, which was partly known. Attempts are also made to find isomeric states which are in general, abundant in the trans-Lead nuclear region\cite{mar,dra,dka}. The shears band, observed in $^{204}$At earlier, has been evaluated in more detail and established to be in reasonable agreement with its mechanism of formation on the basis of the shears mechanism with principal axis cranking (SPAC) model. 

%Your text comes here. Separate text sections with

\section{Experiment Details and Data Analysis}
\begin{figure}[htbp]
%\vskip -1.5cm
\hskip -0.5cm
%\begin{center}
\includegraphics[scale=.48,angle=0]{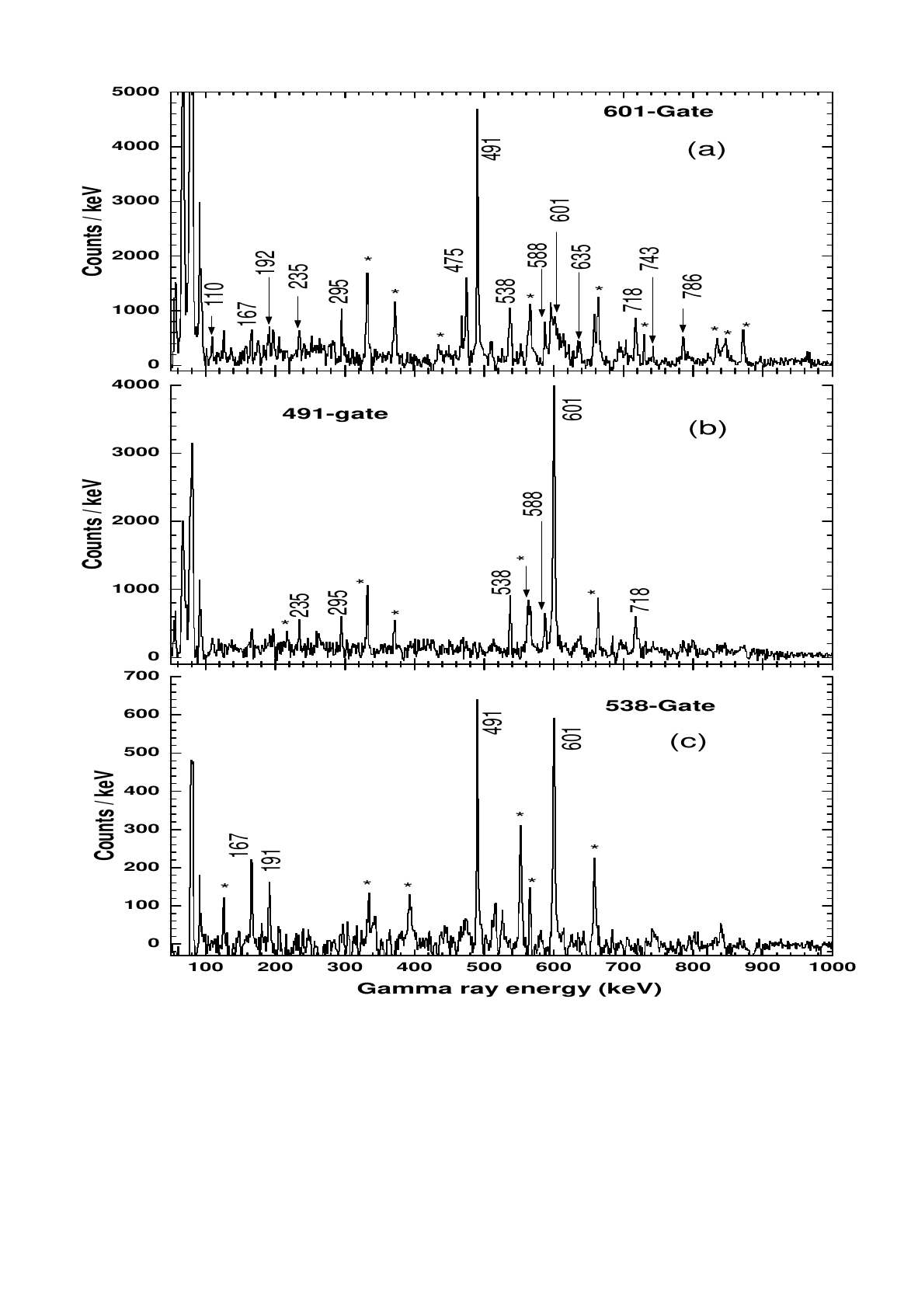} % Fig1
%\end{center}
\vskip -1.5cm
\caption{A few relevant gated spectra of $^{204}$At manifesting mainly the lower part of the level scheme. Majority of the contamination lines are indicated by $*$. See Table~\ref{Tb1} for more details. (a) 601-keV gated spectrum, (b) 491-keV gated spectrum, and (c) 538-keV gated spectrum.}
\label{g601}
%\end{center}
\end{figure}

\begin{figure}[htbp]
%\vskip 6.0cm
\hskip -0.5cm
%\begin{center}
\includegraphics[scale=0.48,angle=0]{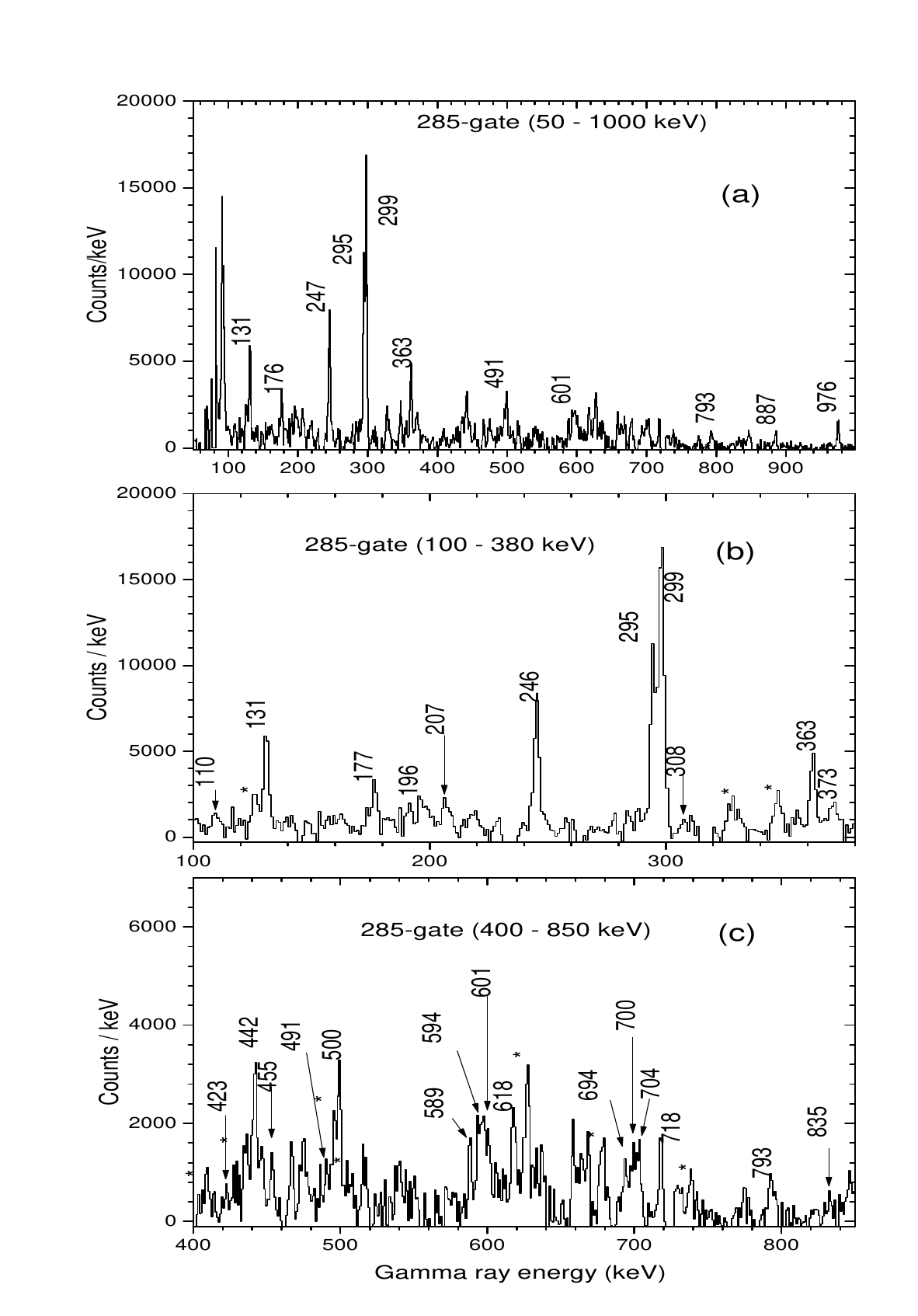} % Fig2
%\end{center}
%\vskip -01.5cm
\caption{(a) Representative spectrum with gate on 285-keV in-band transition belonging to the $M1$-band manifesting mainly the $M1$-band transitions and transitions in coincidence. A few weak crossover $E2$ transitions and the transitions linking the $M1$ band to the ground state are indicated in the expanded plots of the spectra shown in (b) and (c). Major contamination lines are indicated by $*$. See Table~\ref{Tb3} for details.}
\label{285M1}
%\end{center}
\end{figure}

\begin{figure}[htbp]
%\vskip 6.0cm
\hskip -0.5cm
%\begin{center}
\includegraphics[scale=0.48,angle=0]{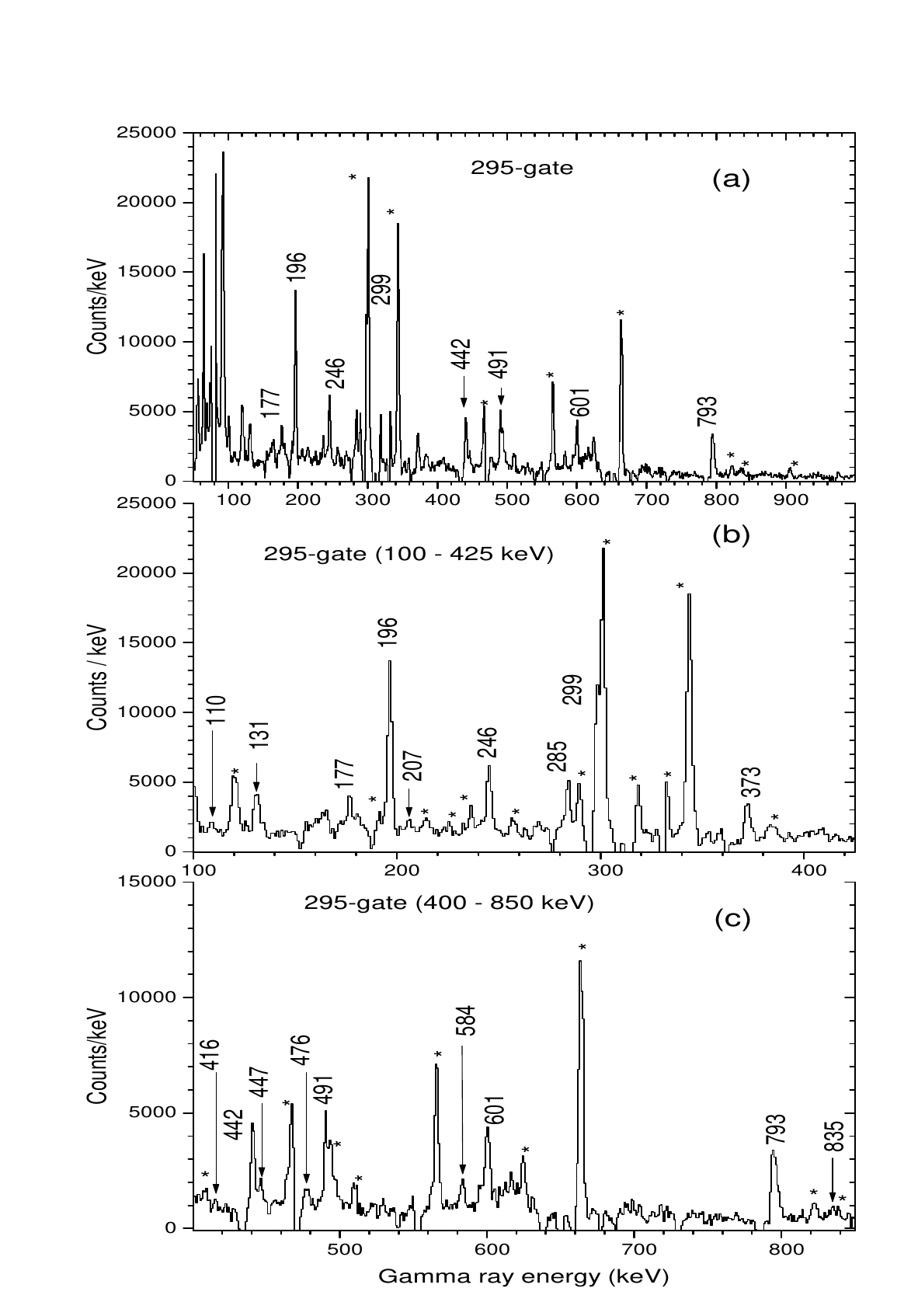} % Fig2
%\end{center}
%\vskip -01.5cm
\caption{Gated spectra similar to those in the Fig.~\ref{285M1}, with the difference that the gate was set on the 295-keV in-band transition belonging to the $M1$-band. Major contamination lines are indicated by $*$. See Table~\ref{Tb3} for details.}
\label{295M1}
%\end{center}
\end{figure}

%\label{sc:expt}
High-spin states of proton rich Astatine nuclei ($^{204-206}$At) were produced by bombarding a $5.0$ mg.cm$^{-2}$ self-supporting Gold ($99.95 \%$ purity) target with $^{12}$C beam at 75 MeV and 65 MeV, provided by the BARC-TIFR Pelletron Linac facility located at the Tata Institute of Fundamental Research (TIFR), Mumbai. Estimation of formation cross sections of the evaporation residues (ER) and the fission yield were done using the code PACE-IV\cite{pace}. Based on these calculations, $\sim 20-25~\%$ of the fusion products at these bombarding energies are estimated to undergo fission. The recoil velocity of the compound nucleus was $ \leq 1 \%$ of \textit{c}. The nucleus $^{204}$At was populated through the $5n \gamma$ evaporation channel, whose estimated yield at 75 MeV was about $ 20 \% $ of the total ER cross-section. However, the yield of excited $^{205}$At nuclei, produced through the $4n \gamma$ evaporation channel, was $\sim 40 - 50\%$ at 75 MeV, which had caused contaminating transitions. The de-exciting prompt $\gamma$-rays were detected using the Compton suppressed HPGe Clover detector array (INGA)\cite{murli} surrounding the target. In our experimental set-up, INGA was configured with fifteen detectors, arranged in a spherical geometry, with three detectors at $157^{\circ}$, four at $90^{\circ}$, and two detectors each at $40^{\circ}$, $65^{\circ}$, $115^{\circ}$ and $140^{\circ}$ with respect to the beam direction. This configuration of detectors was used for angle dependent direction correlation of oriented states (DCO) and polarization direction correlation of oriented states (PDCO) measurements. The Clover detectors were calibrated for $ \gamma$-ray energies and relative efficiencies by using $ ^{133}$Ba and $ ^{152} $Eu radioactive sources by placing them at the target position of the array. Most of the analyses reported here were performed with the data collected at 75 MeV beam energy. We used the data collected at 65 MeV beam energy to confirm the contamination lines from $^{205}$At, since $^{204}$At would not be co-produced at this energy. Approximately 80 million 2-fold, 30 million 3-fold and 8 million 4-fold coincidences were recorded during the experiment at 75 MeV beam energy. Clover coincidence events with time stamps were collected by a fast digital data acquisition (DDAQ) system, based on Pixie-16 modules of XIA LLC \cite{XIA}. A time window of 400 ns was set for this coincidence between the fast triggers of individual channels. Time stamps from a real time clock (RTC), with dwell time of 10~ns, was used for the purpose. The time-stamped data from different Pixie modules were merged to a single data stream in the off-line mode. The data sorting routine Multi pARameter time-stamped based COincidence Search program (MARCOS) developed at the Tata Institute of Fundamental Research (TIFR) \cite{palit,palit12} was used to generate various $\gamma-\gamma$ matrices and $\gamma-\gamma-\gamma$ cubes in a RADWARE and INGASORT \cite{rad,bhow} compatible format for further analysis. The latter program is basically an analysis program for multi-Clover experiments, capable of handling data collected by up to 256 ADC channels, and projected on to 256 spectra at a time using single-and multi-fold coincidences. Angle dependent asymmetric $\gamma-\gamma$ DCO matrices and crystal orientation dependent polarization matrices were constructed using MARCOS. 
The coincidence and the intensity relationships of the $\gamma$-rays were taken into consideration to construct the level scheme of $^{204}$At. Single and double gates were set on the known $\gamma$-ray transitions in the $\gamma-\gamma$ matrices and $\gamma-\gamma - \gamma$ cube for determining the coincidence relations. A few examples of gated $\gamma$-ray spectra from the $\gamma-\gamma$ matrix are shown in the Figures~\ref{g601},~\ref{285M1} and \ref{295M1}. The relevance of these spectra in the level scheme will be discussed in the next section (Sec.~\ref{expt}). 

% Additional text 20/11/21:
Quite a few contamination lines were found to be present in the spectra. Most of these $\gamma$-rays belong to the nuclei produced either by fusion evaporation reactions (ER) or neutron stripping / pickup transfer (Tr) reactions, and they appear because of the coincidence relationship with the gating transitions. For example, the 331 keV line in 601-keV gated spectrum is attributed to excited $^{205}$Po nuclei, produced by the ER process\cite{Fan3}. The same or extremely close 332-keV $\gamma$-rays observed in the 491-keV gated spectrum (see Fig.~\ref{g601}), however, belong to $^{205}$At\cite{sjo}. Because of larger Doppler spread of the $\gamma$-rays produced in the fission channels, large background was observed in the spectra. No overlapping coincident $\gamma$-ray transitions belonging to the relevant fission channel could be identified. A tentative list of contamination lines in the gated spectra (Figs.~\ref{g601}, \ref{285M1} and \ref{295M1}), along with the identified transitions in the nuclei are shown in the Tables~\ref{Tb1} and \ref{Tb3}.
%end of additional text 20/11/21
 
The multipolarities and the electromagnetic character of the observed $ \gamma$-ray transitions for assigning the spin-parity of the levels, have been determined using the DCO\cite{krane,kram89} and the PDCO ratios \cite{dros96,star99,pol92,jone95}. 
%\vskip .5cm
%
%\vskip .5cm
For the DCO ratio analysis, the coincidence events were sorted into an asymmetric matrix with data from the $90^{\circ}$ detectors and from the $157^{\circ}$ detectors. The ratio of $\gamma$-ray intensities, obtained by setting gates on known stretched $E2$ or $M1$ transitions on the two axes of that matrix, were calculated through the following equation:

\beq
\nonumber
R_{DCO} = 
\frac{  I_{\gamma_{2}} ~ observed~at ~ 157^{\circ},~gated~by~ \gamma_{1}~at~90^{\circ}}{ I_{\gamma_{2}}~observed~at~90^{\circ},~gated~by~
\gamma_{1}~at~157^{\circ}}\,, 
\eeq

where, $I_{\gamma_1(\gamma_2)}$ is the number of counts within a particular peak of the detected $\gamma$-ray. The mixing ratio $\delta$ and the theoretical $ R_{DCO} $ were calculated using the software code ANGCOR \cite{angcor} taking the spin alignment parameter $\sigma/J = 0.3$. Theoretically, for a stretched transition, the value of $ R_{DCO} $ would be close to unity for the same multipolarity of $ \gamma_1 $  and $ \gamma_2 $. For different multipolarities and mixed transitions, the values of $ R_{DCO} $ depend on the detector angles, mixing ratio ($ \delta $) and the width of the substate population $(\sigma/J)$. The validity of the $ R_{DCO} $ measurements was checked with the known strong transitions in $^{205}$At \cite{sjo,dav} and with the calculated values. For example, in the present geometry, the measured value of $ R_{DCO} $ for a known 468-keV dipole transition in $^{205}$At, gated by a stretched quadrupole transition, {\em viz.} the 664-keV ground state transition, is $ \approx$ 0.54, while for a quadrupole transition gated by a pure dipole, the calculated value is $ \approx $ 1.85. Moreover the value of $ R_{DCO} $ for a stretched quadrupole transition (566 keV), gated by the 664-keV ground state transition, is $ \approx 1.01$. These results are consistent with the theoretical values.

The linear polarisation of the $\gamma$-ray transitions has been experimentally measured by the use of the Clover detectors as Compton polarimeters, utilising the direction of the scattered radiation to determine the electric or magnetic nature of the transition \cite{pol92,jone95}. The polarization asymmetry ($\Delta_{PDCO}$) of the Compton scattered photons inside the detector medium, is defined as:
\beq
\nonumber
\Delta_{PDCO} = \frac{a(E_{\gamma})N_{\perp} - N_{\parallel}}{a(E_{\gamma})N_{\perp} + N_{\parallel}}\,,
\eeq
where, $N_{\perp}$ and $ N_{\parallel}$ are the number of $\gamma$-rays of a particular energy, Compton  scattered in the planes perpendicular and parallel to the reaction plane. $a(E_{\gamma})$ is the correctional term due to the geometrical asymmetry of the Clover crystals in the array set up, and is defined as: 
\beq
\nonumber
a(E_{\gamma}) = \frac{N_{\parallel}(unpolarised)}{N_{\perp}(unpolarised)}\,.
\eeq
The linear polarization ($P$) of the associated $\gamma$-ray is directly proportional to the polarization asymmetry $\Delta_{PDCO} = P \, Q(E_\gamma)$,  where $Q(E_\gamma)$ is the polarization sensitivity of the Clover detector of the INGA set up. The $Q(E_ \gamma)$ has been demonstrated\cite{Raj} to have very weak dependence on $E_\gamma$ within the expected uncertainty of the measurement system for $E_\gamma \lesssim 1000$~keV, and therefore, $\Delta_{PDCO}$ can be considered as a measure of the linear polarization $P$ of the $\gamma$-rays for identifying the electromagnetic nature of the transitions. 

\begin{figure}[htbp]
%\vskip -0.5cm
\hskip -1.0cm
%\begin{center}
\includegraphics[scale=.35,angle=0]{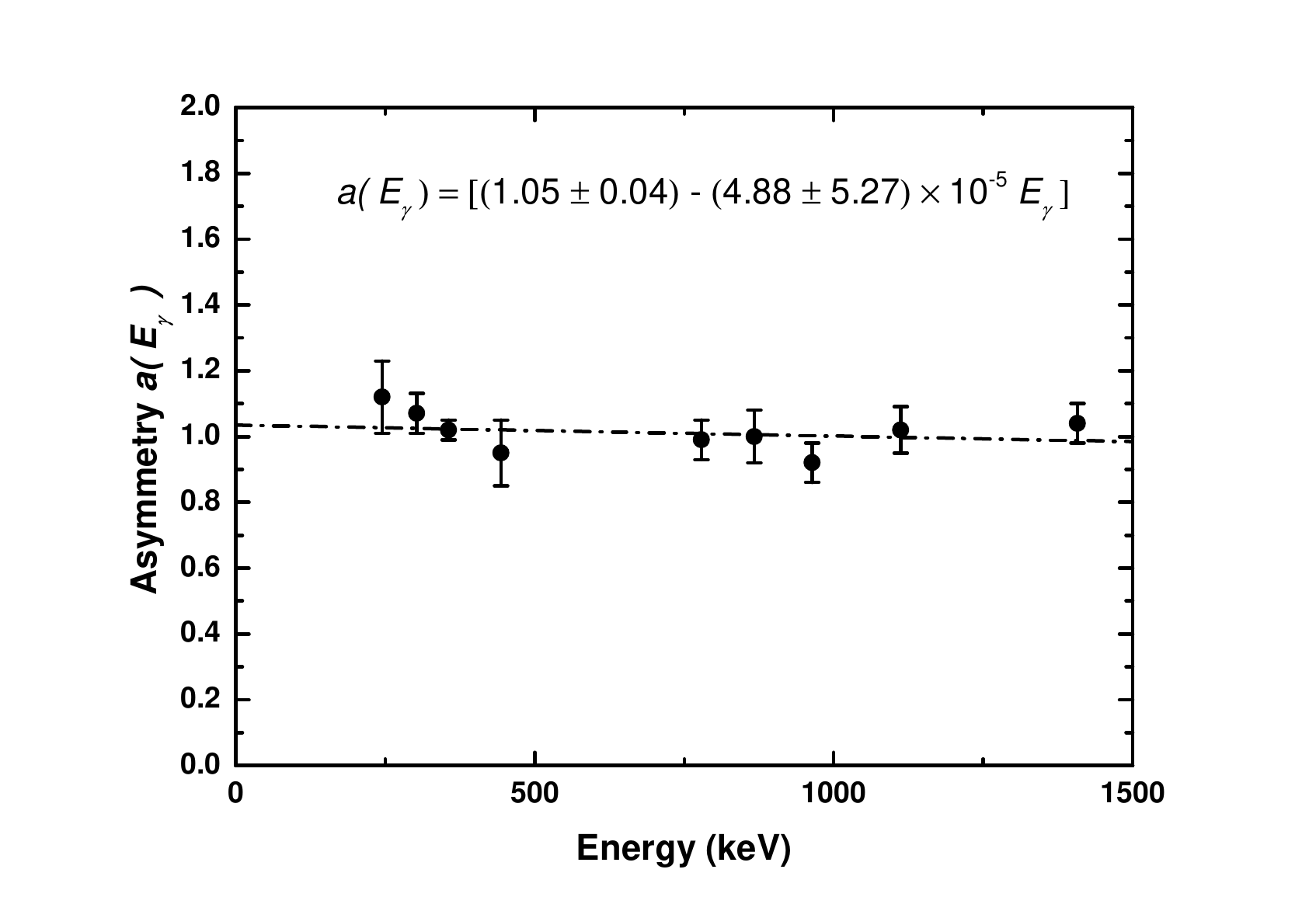} % Fig3
%\end{center}
%\vskip -01.0cm
\caption{The asymmetry correction factor $a(E_{\gamma})$ for the clover
detector placed at $90^{\circ}$ with respect to the beam direction. The dash-dotted
line corresponds to linear fit to the data points.}
\label{asym}
%\end{center}
\end{figure}

The asymmetry correction factor $a(E_{\gamma})$ was evaluated as a function of energy using $^{152}$Eu and $^{133}$Ba sources placed at the target position inside the array and was found to be 1.05(4), over the energy range $300 - 1400$~keV. The measured $a(E_{\gamma})$ values are shown in Fig.~\ref{asym}. 

%\vskip -2cm

By using the fitted parameter $a(E_{\gamma})$, the PDCO of the $\gamma$-rays in $^{204}$At have been determined. A positive value of PDCO indicates an electric-type transition whereas a negative value favors magnetic-type transition. PDCO values could not be measured for the low energy and the weak transitions. The low-energy cutoff for the PDCO measurements was about 200~keV in this work. Like DCO, the validity of the method of the PDCO measurements was also confirmed from the known transitions in $^{205}$At. The experimental asymmetry, for the transitions of interest, was evaluated from the two $ \gamma-\gamma $ matrices. One axis of both the matrices corresponds to the event recorded in any clover detector while the other axis contains the coincident scattered events inside the clover detectors placed at $ 90^{\circ}$ with respect to the beam axis, in a direction perpendicular or parallel to the emission plane. 
\begin{table}[htbp]
\caption{Major contamination lines arising from overlap with the 601, 491 and 538-keV gating transitions (see Fig.~\ref{g601}). $\Delta E$ window of 4 keV was used to search for overlapping coincident transitions. Abbreviation used: ER for evaporation residue, TR for nucleon transfer.}
%\vskip 0.2cm 
%\begin{center}
%\begin{ruledtabular}
\begin{threeparttable}
\begin{tabular}{lllll} \hline
Gate on & Contaminant & Nucleus & Reaction & Ref. \\ 
$E_{\gamma 1}$ (keV) & $E_{\gamma 2}$ (keV) & & & \\ \hline
\\
%\hline
601 & 320 & $^{205}$Bi & ER \tnote{1} & \cite{Alp} \\
    & 331 & $^{205}$Po & ER & \cite{Fan3} \\
    & 372 & $^{205}$Po & ER & \cite{Fan3} \\
    & 433 & $^{206}$Po & ER & \cite{Bax} \\
    & 568 & $^{206}$Po & ER & \cite{Bax} \\
    & 663 & Unplaced & & \\
    & 729 & $^{206}$Po & ER & \cite{Bax} \\
    & 834 & $^{205}$Bi & ER\tnote{1} & \cite{Alp} \\
    & 847 & $^{205}$Po & ER & \cite{Fan3} \\
    & 872 & $^{205}$Bi & ER\tnote{1} & \cite{Alp} \\
491 & 216 & $^{198}$Au & TR & \cite{198Au} \\
    & 332 & $^{205}$At & ER & \cite{sjo} \\
    & 371 & $^{205}$At & ER & \cite{sjo} \\
    & 567 & $^{198}$Au & TR & \cite{198Au} \\
    & 665 & $^{199}$Au & TR & \cite{199Au} \\
538 & 126 & $^{205}$At & ER & \cite{dav} \\
    & 332 & $^{205}$At & ER & \cite{dav} \\
    & 396 & $^{205}$At & ER & \cite{dav} \\
    & 553 & $^{205}$At & ER & \cite{dav} \\
    & 566 & $^{205}$At & ER & \cite{dav} \\
    & 659 & $^{205}$At & ER & \cite{dav} \\
\hline
%\hline 
\end{tabular}
\begin{tablenotes}
\item[1] Produced from $^{205}$Po ER by electron capture ($T_{1/2} = 1.7$~Hrs). 
\end{tablenotes}
\end{threeparttable}
%\end{ruledtabular}
\label{Tb1}
%\end{center}
\end{table}

\begin{table}[htbp]
\caption{Major contamination lines arising from overlap with the 285 and 295-keV gating transitions (see Figs.~\ref{285M1} \& \ref{295M1}). $\Delta E$ window of 4 keV was used to search for overlapping coincident transitions.}
%\vskip 0.2cm 
%\begin{center}
%\begin{ruledtabular}
%\begin{threeparttable}
\begin{tabular}{lllll}
\hline
Gate on & Contaminant & Nucleus & Reaction & Ref. \\ 
$E_{\gamma 1}$ (keV) & $E_{\gamma 2}$ (keV) & & & \\ \hline
\\
%\hline
285 & 124 & $^{198}$Au & Tr & \cite{198Au} \\
    & 328 & $^{198}$Au & Tr & \cite{198Au} \\
    & 348 & $^{198}$Au & Tr & \cite{198Au} \\
    & 408 & $^{198}$Au & Tr & \cite{198Au} \\
    & 433 & $^{198}$Au & Tr & \cite{198Au} \\
    & 498 & $^{198}$Au & Tr & \cite{198Au} \\
    & 516 & $^{204}$Po & ER & \cite{204Po} \\
    & 626 & $^{198}$Au & Tr & \cite{198Au} \\
    & 679 & Unplaced & & \\
    & 739 & Unplaced & & \\
    & 847 & Unplaced & & \\
295 & 123 & $^{198}$Au & Tr & \cite{198Au} \\
    & 193 & $^{198}$Au & Tr & \cite{198Au} \\
    & 215 & $^{198}$Au & Tr & \cite{198Au} \\
    & 225 & $^{198}$Au & Tr & \cite{198Au} \\
    & 236 & $^{198}$Au & Tr & \cite{198Au} \\
    & 257 & $^{198}$Au & Tr & \cite{198Au} \\
    & 289 & $^{198}$Au & Tr & \cite{198Au} \\
    & 301 & $^{198}$Au & Tr & \cite{198Au} \\
    & 319 & $^{198}$Au & Tr & \cite{198Au} \\
    & 333 & $^{198}$Au & Tr & \cite{198Au} \\
    & 344 & $^{198}$Au & Tr & \cite{198Au} \\
    & 386 & $^{198}$Au & Tr & \cite{198Au} \\
    & 409 & Unplaced & & \\
    & 468 & Unplaced & & \\
    & 495 & Unplaced & & \\
    & 508 & Unplaced & & \\
    & 566 & $^{198}$Au & Tr & \cite{198Au} \\
    & 616 & $^{204}$Po & ER & \cite{204Po} \\
    & 626 & $^{198}$Au & Tr & \cite{198Au} \\
    & 664 & Unplaced & & \\
    & 822 & $^{204}$Po & ER & \cite{204Po} \\
    & 839 & $^{198}$Au & Tr & \cite{198Au} \\
    & 905 & Unplaced && \\
%\hline \hline
\hline 
\end{tabular}
%\end{threeparttable}
%\end{ruledtabular}
\label{Tb3}
%\end{center}
\end{table}

\section{Experimental Results}
\label{expt}
Gamma-ray spectroscopy of $^{204}$At was attempted with the aims to confirm and extend the previously reported results\cite{hart}, and to investigate in detail a possible magnetic rotational (MR) band within the framework of shell model approximation in this doubly odd At isotope. The ground state of $^{204}$At has spin-parity of $7^{+}$, and is known to be of $\pi(1h_{9/2}) \otimes \nu(2f_{5/2}^{-1})$ configuration outside the $^{202}$Po core. However, the 587-keV $10^{-}$ first excited state is a $j$-forbidden isomeric state with $ \sim$ 100 ms lifetime, resulting from the $\pi(1h_{9/2}) \otimes \nu(1i_{13/2}^{-1})$ configuration as identified earlier\cite{gip}. The main sequence of transitions including 601-, 491-, 717-keV $\gamma$-rays and the $ \Delta I $ = 1 sequence of transitions involving 131-, 296-, 299- and 246-keV $\gamma$-rays were identified by the previous workers\cite{hart}. All the previously observed $\gamma$-ray transitions were observed in the present experiment. Majority of the transitions in $^{204}$At are reported here for the first time.

\begin{figure*}
%\vskip 6.0cm
%\hskip -4.0cm
\begin{center}
\includegraphics[scale=0.85,angle=0,trim={2.5cm 2.5cm 2.5cm 2.5cm},clip]{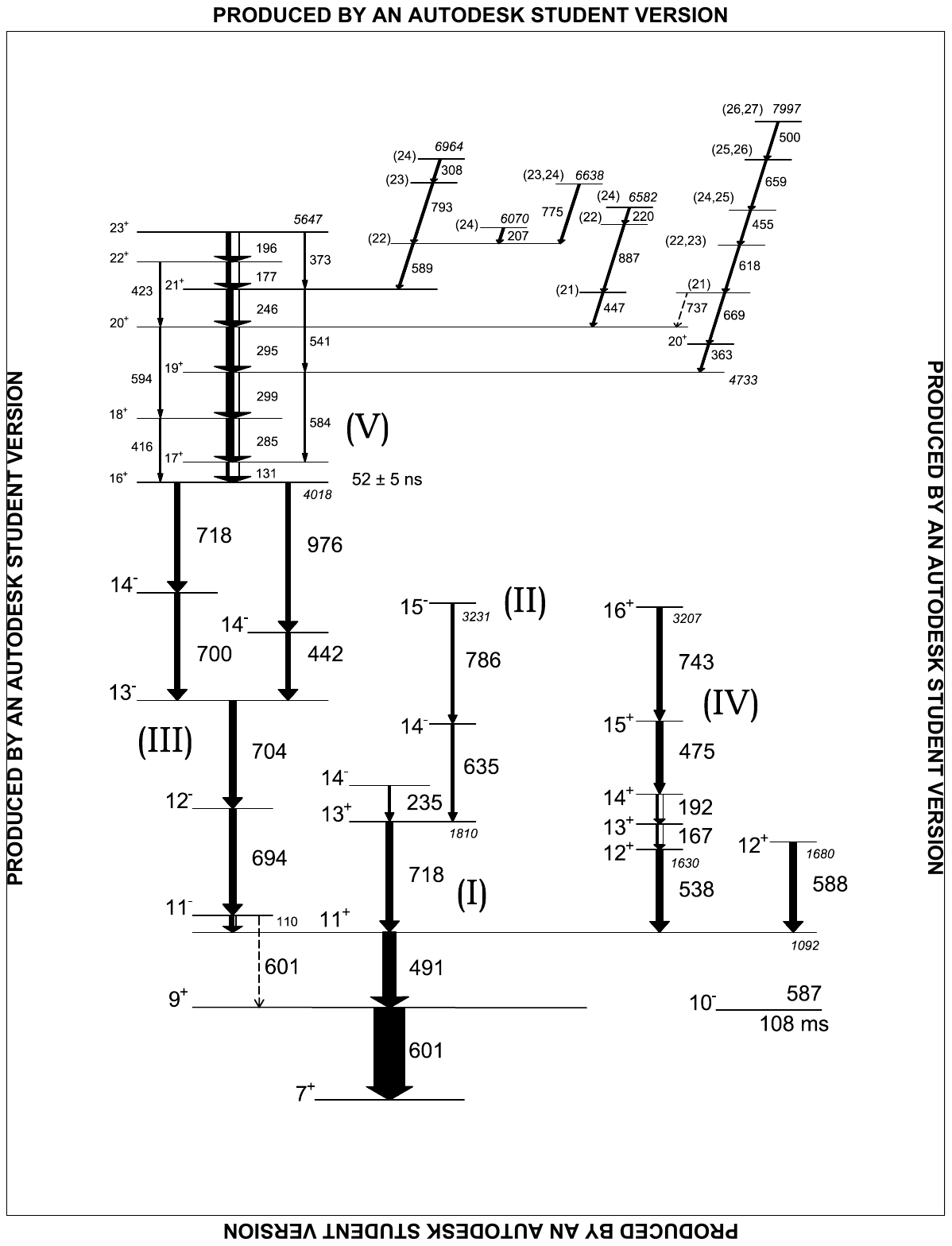} % Fig4
\end{center}
%\vskip -08.2cm
\caption{Proposed level scheme of $^{204}$At obtained from the present work, except for the 108~ms $10^-$ isomer at 587 keV\cite{gip}. Transition energies are rounded off to the nearest whole number values in keV. Fraction of internal conversion contribution to the total intensity is indicated by the width of white patches of the arrows. Part of the level scheme showing transitions feeding the dipole band of sequence V was not drawn to scale.}
\label{level}
%\end{center}
\end{figure*}
%\vskip .5cm

\begin{figure}[htbp]
%\vskip 6.0cm
%\hskip -4.0cm
\begin{center}
\includegraphics[scale=.35,angle=0]{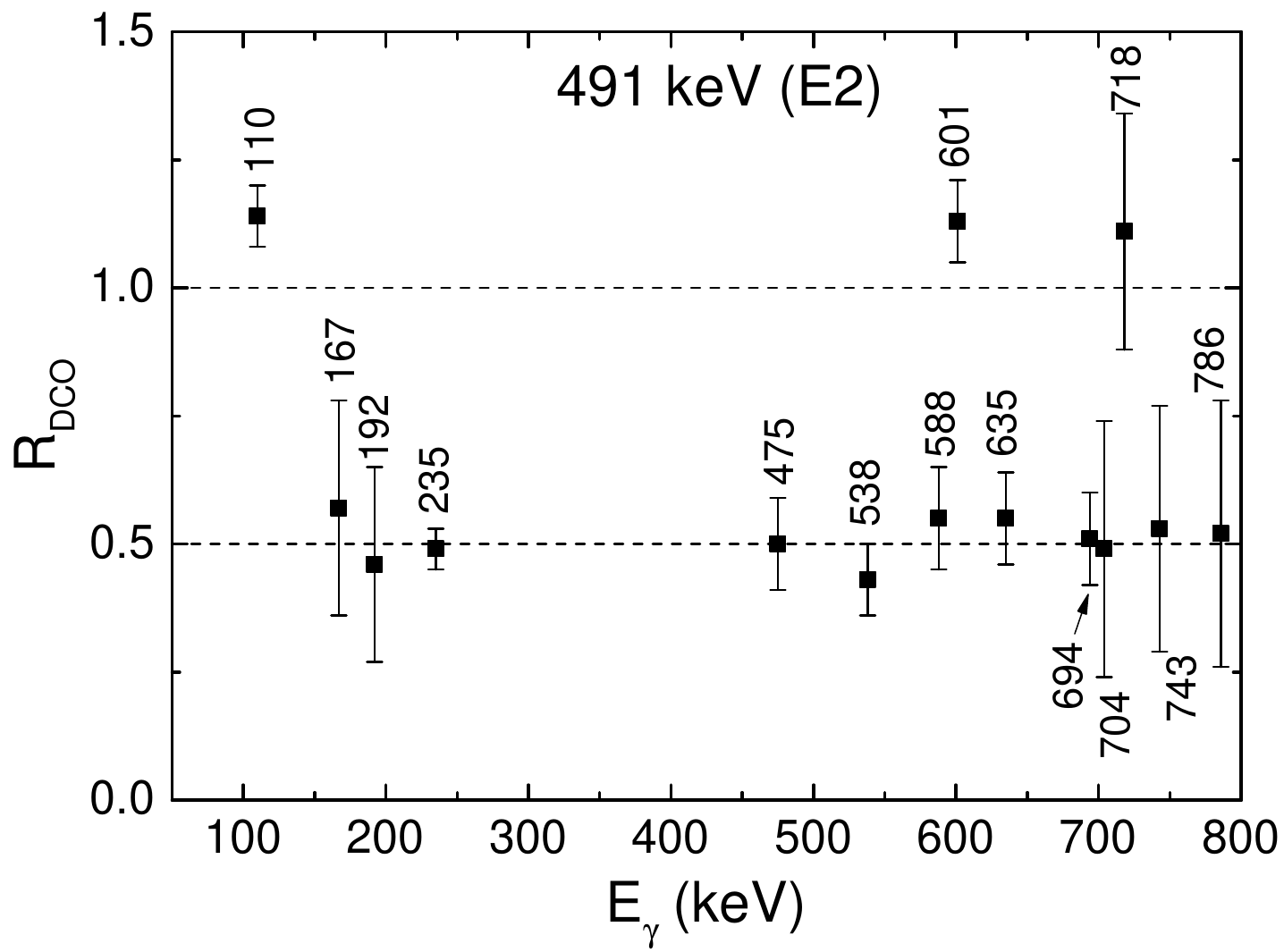} % Fig5
\end{center}
%\vskip -08.2cm
\caption{Plots of DCO ratios for a few transitions with respect to the 491-keV $E2$ transitions in $^{204}$At, except for the 110-keV transition for which, the ratio was estimated with respect to 601-keV.}
\label{dco1}
%\end{center}
\end{figure}

\begin{figure}[htbp]
%\vskip 6.0cm
%\hskip -4.0cm
\begin{center}
\includegraphics[scale=.30,angle=0]{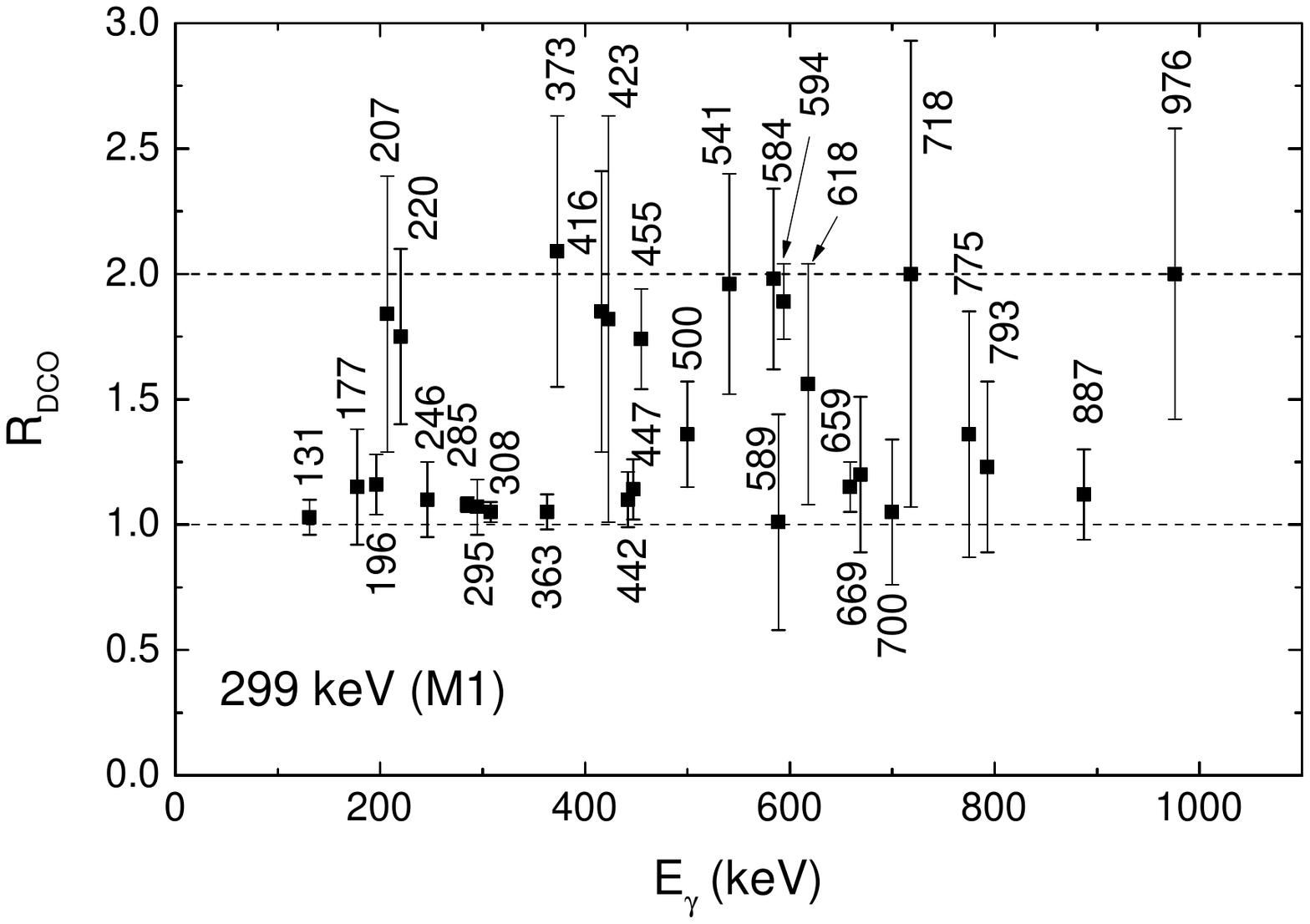} % Fig6
\end{center}
%\vskip -08.2cm
\caption{Plots of DCO ratios for a few transitions with respect to the 299-keV $M1$ transitions in $^{204}$At, except for the 423, 541, 594, 584, and 659-keV transitions for which, DCO ratios are measured with respect to 285- or 295-keV transition of the dipole band. See Table~\ref{Tb2} for details.}
\label{dco2}
%\end{center}
\end{figure}

\begin{figure}[htbp]
%\vskip 6.0cm
%\hskip -4.0cm
\begin{center}
\includegraphics[scale=.40,angle=0]{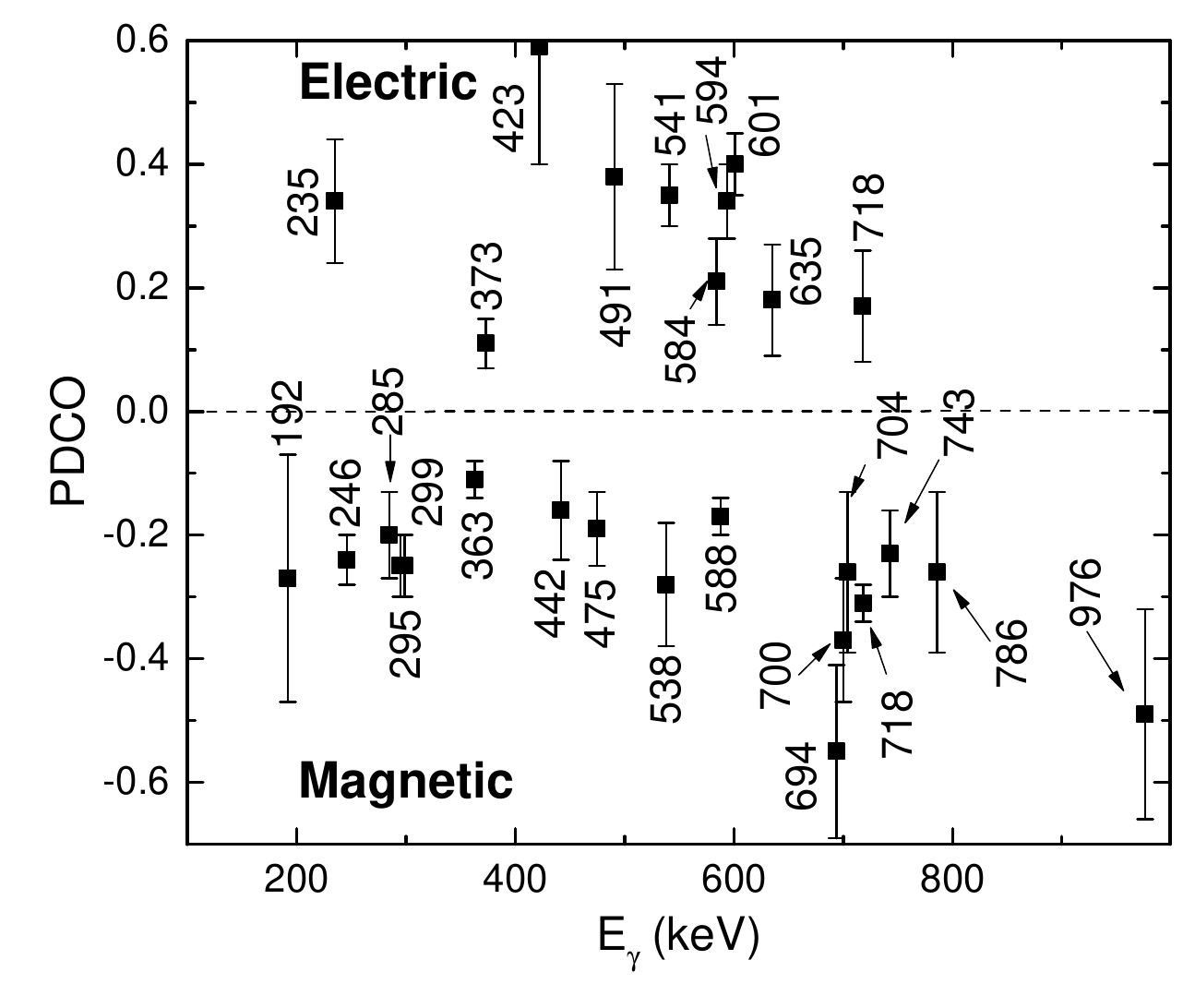} % Fig7
\end{center}
%\vskip -08.2cm
\caption{Plots of PDCO ratios for a few transitions in $^{204}$At. PDCO for both the 718-keV transitions are shown. $\Delta_{\rm PDCO} >0$ is for the sequence I transition, while the other one belongs to the sequence III.}
\label{pdco}
%\end{center}
\end{figure}

The proposed level scheme of $^{204}$At (Fig.~\ref{level}), obtained from the present experiment, is established using the $\gamma-\gamma $ coincidence information, relative intensity, and multipolarity information from the $R_{DCO} $, and linear polarization measurements. The width of the arrows in the figure indicates the total decay intensity (sum total of the $\gamma$-ray intensity and the expected contribution arising from internal conversion, estimated from the Ref.~\cite{Kib}). The existing level scheme has been modified with better statistics and a few new branches of transitions identified in this work. From the prompt $\gamma-\gamma $ matrices, constructed from the online data taken at 75 MeV beam energy and by gating on the 601-keV ground state transition, the intense 491-keV transition (Fig.~\ref{g601}~(a)), and also on the Astatine $X$-rays, a significant number of new low lying transitions were observed. The sequences of transitions in the level scheme are indicated by the roman numerals (I - V). A few new transitions linking the $M1$ band (sequence V) with the branch populating the 601-keV  $9^+$ level could also be observed and fitted into the level sequence. Further investigations into the cross correlation of the transitions reveal the main yrast sequence I. The $\gamma$-ray spectra, obtained by gating on the known transitions of the $ \Delta I$ = 1 band (Figures~\ref{285M1} and \ref{295M1}), reveal six new cross-over transitions and several other new transitions. However, these were not found in coincidence with any of the known $^{205}$At or $^{203}$At gamma rays. DCO ratios and polarization asymmetry results were used to estimate the multipolarity and to ascertain the electric or magnetic nature of the transitions (Figs.~{\ref{dco1}, \ref{dco2}, and \ref{pdco}}). The deduced excitation energy, associated relative intensity of the observed $\gamma$-ray transitions, expected spin-parity of the excited levels based on our DCO ratio and PDCO measurements, together with other relevant information regarding their placement in the proposed level scheme of $^{204}$At, are summarized in the Table~\ref{Tb2}. A few transitions in the spectrum, which could not be placed in the level scheme due to inconsistency in intensity, spin-parity and coincidence conditions, are also mentioned in the table. For a few transitions, ambiguity arises regarding coincidence-intensity as we could not avoid contamination from the nearby strong $\gamma$-rays belonging to $^{205}$At. The major contamination lines, which were observed in coincidence with the gating transitions are indicated in the Figures~\ref{g601}, \ref{285M1} and \ref{295M1}, and are also enlisted in the Tables~\ref{Tb1} and \ref{Tb3}.  

%\vskip .5cm

%\vskip .5cm

%\vskip .5cm

The level scheme of $^{204}$At has been extended to an excitation energy of $\lesssim 8.0$~MeV and $\sim 25\,\hbar$, and is a much improved one compared to the previously known level scheme reported in Ref.~\cite{hart}. The relative $\gamma$-ray intensities given in the table were measured and normalised using two different coincidence gates. In the first case, the gate was set on the 601-keV ground state transition. In the second case, intensities of the $\gamma$-ray transitions from the states  above the $13^{-}$ state (the sequence V transitions and others feeding them) are normalized to the intensity of the 285-keV transition. Because of the existence of low lying isomers with lifetimes (ns to ms), and also due to the large internal conversion of some of the levels, the intensity balance across the levels could be done with some approximations. For most of the transitions, relative intensities are corrected by taking into account the estimated electron conversion coefficients\cite{Kib}. The observed $\gamma$-rays and their relative intensities match qualitatively with those obtained by the previous workers\cite{hart}. Based on the intensity correlations obtained from the gated spectra, the DCO and the PDCO ratios, the level scheme for $^{204}$At is clearly established as shown in Fig.~\ref{level}. 

There is more than one sequence of transitions populating the 1092-keV $11^+$ level. Out of these, sequence II passing through 786-, 635-, and 718-keV and the sequence IV consisting of 743-, 475-, 192-, 167-, and 538-keV transitions were observed. The major sequence of transitions (III) passing through 110-, 694- and 704-keV, extends via the sequence of $M1$ band transitions (V) all the way to $\sim 7$~MeV of excitation energy. In all the sequences, ordering of the transitions were cross-checked by intensity correlations and also by reverse gating as far as possible. About 30 new transitions, over and above those observed by Hartley et al.~\cite{hart}, were found, as listed in the Table~\ref{Tb2}. All these lines were found in coincidence with the Astatine $X$-rays. A few relevant gated spectra are shown in the Figures~\ref{g601},~\ref{285M1} and \ref{295M1}. The 1680-keV level decays to 1092-keV level via 588-keV $\gamma$-rays (reported as 589-keV in the earlier work). From the measured $ R_{DCO} $ and polarization asymmetry, it is concluded that the 588-keV transition has a pure dipole character with no change in parity. We were unable to observe any other new transitions to extend this branch beyond the 1680-keV level. 

In the previous work by Hartley et al.~\cite{hart}, the nature of the 601-keV ground state transition was not clear and tentatively assigned as dipole. In the present work, however, this assignment is modified to electric quadrupole in nature based on DCO and PDCO measurements (see Table~\ref{Tb2}). The DCO ratio for this $\gamma$-ray has been measured by gating on the 491-keV $\gamma $-ray, which was known to be a quadrupole $(E2)$ transition on the basis of angular distribution analysis\cite{hart}. 
The 718-keV transition is confirmed in the present work as $ E2$ transition based on the DCO and the PDCO ratios in agreement with the earlier results. Intensity correlation also confirms the order of placement along the cascade.

The 2347-keV level, observed in the previous work\cite{hart}, was found to decay through the 537- and 717-keV transitions via the $491 - 601$-keV sequence. In the present work, this assignment was re-examined and modified as: $ 537.9(12)$ and $717.8(14)$~keV, and henceforth designated as 538- and 718-keV $\gamma$-rays respectively. It was observed that the 538-keV transition, though passing through the $491-601$-keV sequence, was not in coincidence with the 718-keV $\gamma$-ray (Fig.~\ref{g601}~(a)). Therefore, the 538-keV transition is placed in the sequence IV depopulating the 1630-keV level. The 718-keV transition was identified as $E2$, whereas the 538-keV transition was found to be $M1$ (see Table~\ref{Tb2}). Newly observed 743-, 475-, 192-, and 167-keV $\gamma $-rays have been placed above the 1630-keV level in the sequence IV based on the corresponding coincidence and intensity relations (see Fig.~\ref{g601}~(a)), and were found to be $M1$ (see Table~\ref{Tb2}) based on the DCO and the PDCO values, except for the 167-keV transition, where polarization could not be confirmed. However, we could infer $M1$ nature of the transition from the DCO ratio and the sequence of transitions.
The 786-keV $M1$  and 635-keV $E1$ transitions are placed in the sequence II above the 1810-keV level. A new transition with $E_\gamma=235$~keV, 
depopulating through sequence I and parallel to the sequence II, was identified as $E1$. This was observed only in the 718-, 491- and 601-keV gate, but not in coincidence with any other transition of nearby parallel sequences.

A new sequence III including 704-, 694- and 110-keV transitions was observed as feeding the 1092-keV 11+ level. The intensities of the three transitions were found to be the same within the respective uncertainties. Since the above observation does not justify the sequence of their placements, we have obtained the intensities by gating on the 285-keV transition, which feeds the sequence from above.  Intensities normalized with respect to 285-keV for the 704-, 694- and 110-keV transitions were obtained as: $262\pm 37,\, 85\pm 21,\, 134\pm 25$ respectively. Intensity mismatch for the three transitions belonging to the sequence III was observed when reverse gating was done to extract the intensities. While this justifies the placement sequence of 704- and 694-keV transitions, the 110-keV transition displays a bit of excess intensity. This possibly indicates additional feeding to the 1202-keV level. We could not find any other transition, except for a weak 601-keV transition present in 601-keV gate (see Fig.\ref{g601}), which may be placed as shown in the level scheme by dotted line. This would ensure the placement of the 110-keV transition in the sequence. However, the DCO and PDCO measurements on the weak 601-keV transition were not possible due to the presence of 601-keV ground state transition.
For the 110-keV transition, only DCO ratio estimation could be done. Based on our measured DCO ratio of $1.14(6)$, gated by the 601-keV $E2$ transition, $ \Delta J = 0 $, $ E1$ ($ 11^{-} \rightarrow 11^{+} $) character was assigned. Based on this result, ($ J^{\pi} $) of $ 11^{-} $ has been assigned to the 1202-keV level. The configuration of the initial and the final states are different as highlighted in the interpretation section. Intensity mismatch for the transition belonging to the sequence III was observed, which may be attributed partly to significant internal conversion $(\alpha \sim 0.4$). Weisskopf estimate for lifetime of this $E1$ transition was $2 \times 10^{-13}\, {\rm sec}$. Despite our isomer search within the limitations of the experiment, no measurable lifetime was found.

Several parallel sequences of transitions were observed between the $13^{-} $ and $ 16^{+} $ states above the sequence III. Two such sequences are indicated in the level scheme of Fig.~\ref{level}. 
The intensities of the 976- and the 442-keV transitions were almost equal and so were those of the 718-keV and the 700-keV transitions. While building up the tentative level scheme of $^{204}$At, a few transitions, such as the 442-keV transition, were reported to have coincidence relationship with $\gamma$-rays belonging to the $M1$ band, although these could not be placed in the level scheme\cite{hart}. Based on our data and analysis, this was placed between the states $13^{-}$ and $ 14^{-}$, along with other transitions, all of which are in coincidence with $\gamma$-rays belonging to the $M1$ band and the At $X$-rays. 
 
Figures~\ref{285M1} and \ref{295M1} show cascade of prompt $\gamma$-rays of the $M1$ band in the 285-, and 295-keV gated spectra. The band-head energy and spin of the band has been suggested to be of 4018~keV and $16^{+}$, based on energy, spin summing of $\gamma$-rays and multipolarity information. However, any change in the band-head spin or missing levels in between would simply shift the level scheme along the vertical axis.
The relative placement of the $\gamma$-rays in the band is based on their coincidence relations and intensity profiles, taking into account the theoretical total electron conversion coefficients \cite{Kib}. This ordering is the same as reported earlier by Hartley {\em et al}. The spins and parities of the states in this band are assigned from the measured DCO and the PDCO ratios, wherever possible. The $R_{DCO} $ values of these transitions are very close to the expected values for pure dipole transitions. The negative values obtained for the PDCO ratios for the 285-, 299-, 295- and the 246-keV transitions, together with their $R_{DCO} $ values and high $X$-ray yield in coincidence, give clear evidence that they are predominantly $M1$ in nature. 
Since the other $\gamma$-rays like 131-, 177- and 196-keV are also in-band with the $M1$ transitions, and manifested $R_{DCO} $ values expected for pure dipole transitions, we have assigned $M1$ nature for these $\gamma$-rays, although the sign of the polarisation asymmetry could not be determined unambiguously due to small Compton scattering probability for low energy $\gamma $-rays ($E_{\gamma} \lesssim 200$~keV). Indications for most of the crossover $E2$ transitions were found (see Figures~\ref{285M1} and \ref{295M1}) in the present work, which were not observed earlier. 
Measured DCO and the PDCO ratios identified the electric quadrupole nature of these transitions. Observation of crossover $E2$ transitions lead to conclusion beyond doubt about the correct ordering of the dipole transitions along the band. Doppler lineshape analysis to determine the lifetimes and consequent estimation of the transition rates $B(M1)$ and $B(E2)$ could not be done due to limitations in our experimental arrangements. However, the ratio of reduced transition probabilities $B(M1) / B(E2)$ could be extracted from experimental $\gamma$-ray branching ratios of competing $ \Delta I = 2 $ and $ \Delta I = 1 $ transitions \cite{bm1} as: 

\beq
%\nonumber
\frac{B(M1)}{B(E2)} = 0.697 \frac{E_{2}^{5}}{E^{3}_{1}} \frac{1}{1 + \delta^{2}} \frac{I_{\gamma}(\Delta I = 1)}{I_{\gamma}(\Delta I = 2)}.
\label{BM1BE2}
\eeq

$E_{1}$ and $E_{2} $ are the energy of the $M 1 (\Delta I = 1)$ and $E 2 (\Delta I = 2)$ $ \gamma$-ray transitions in MeV respectively. The value of $ \delta $ determines the $E2/M1$ mixing ratio of the $ \Delta I = 1 $, $M1$ transition. The mixing ratio $\delta$, estimated from the DCO ratios for the 285-, 295- and 299-keV $M1$ transitions belonging to the band in sequence V, were found to be $\lesssim 0.02(8)$, indicating negligible mixing of $E2$ for these transitions. Table~\ref{Tb2} also lists the values for the experimental $B(M1) / B(E2)$ ratios where ever possible to identify the crossover $E2$ transitions. These ratios are found to be quite large ($ \geq 30\,\mu_{N}^{2} / e^{2}b^{2}) $, suggesting magnetic rotational nature of the band. 

A few other band-like structures with large number of new transitions are also observed in the present work through conventional $ \gamma-\gamma $ and $\gamma-\gamma-\gamma $ coincidence analysis. The sequence of 363-, 669-, 618-, 455-, 659- and 500-keV transitions were found to be in coincidence with the $M1$ band transitions and the At $ X$-rays. These are placed above the $19^{+}$ states of the $M1$ band accordingly. The intensity profile of each transition was calculated from several alternative gates at 285-keV, 299-keV and 363-keV. Similarly, another two minor sequences of transitions composed of 447-887-220-keV and 589-793-308-keV have been placed above the $20^{+}$ and $21^{+}$ levels of the $M1$ band respectively. The observed $\gamma$-rays belonging to these branches, can be seen in the spectra shown in the Figures~\ref{285M1},~\ref{295M1} and are listed as second group in the Table~\ref{Tb2}. Due to limitations in the counts and contamination in some of the gating transitions, the electromagnetic characters of most of these transitions could not be firmly established. However, $M1$ character of the 363-keV transition (5096 $ \rightarrow $ 4733~keV) was clearly established. Above this 5096-keV level, tentative DCO ratio estimation was possible, which indicates that these are mostly dipole transitions. In the absence of polarization data, we have tentatively assigned the spin of these levels. Similar arguments may be given for other minor branches of weaker transitions for which, tentative spin assignments were done. 

\begin{table*}[htbp]
%\vskip 1.9cm
\caption{Energies ($ E_{\gamma}$), associated relative photon intensities ($ I_{\gamma} $), total intensities including internal conversion ($I_{\it total}$), energies of initial levels ($ E_{i} $), DCO ratios ($ R_{DCO} $), PDCO ratios ($ \Delta_{PDCO} $ ) along with deduced multipolarities of the $ \gamma $-rays in $^{204}$At are given. The spins and parities of initial ($ J^{\pi}_{i} $) and final ($ J^{\pi}_{f} $) levels are also given. Theoretical conversion coefficients (Ref.~\cite{Kib}) are considered. Intensities are normalized relative to the intensity of 601-keV ground state transition in raw spectrum for the first group (low lying transitions) and of the 285-keV transition for the second group (associated with the shears band) of transitions. Unless otherwise specified, the DCO ratios are measured with respect to the 491-keV stretched quadrupole\cite{hart} and the 299-keV stretched dipole transitions respectively for the two groups separated by horizontal line in the table. Corresponding experimental $B(M1)/B(E2)$ for the dipole band are also quoted.}
%\vskip 0.2cm 
%\begin{center}
\begin{threeparttable}
\begin{tabular}{llllllllll}
\hline
$E_\gamma$ & $I_\gamma$ & $I_{total}$ & $ E_{i} $ & $ J^{\pi}_{i} $  & $ J^{\pi}_{f} $  &  $ R_{DCO}$  &  $ \Delta_{PDCO} $  &  Multipolarity  & $B(M1)/B(E2)$\\ 
(keV) &   &  & (keV)  &  &  &  &  & Assigned & ($ \mu_{N}^{2} / e^{2}b^{2} $) \\ \hline 
110.1 & 11.7(10) & 15.9(13)   & 1202 & 11$ ^{-}$  & 11$^{+}$      & 1.14(6)\tnote{1} & & $E1$ \\
166.5 & 15.6(16) & 59.3(58)    & 1797 & 13$ ^{+} $ & 12$ ^{+} $  & 0.57(21)                 &  & $M1$\\
191.6 & 15.6(10) & 45.2(31)     & 1989 & 14$ ^{+} $ & 13$ ^{+} $  & 0.46(19)                 & -0.27(20)  & $M1$\\
234.6 & 15.6(10) & 16.4(11)     & 2045 & 14$ ^{-} $ & 13$ ^{+} $  & 0.49(4)                  & +0.34(10) & $E1$\\ 
475.3 & 75.4(50)\tnote{2} & 87.3(59)\tnote{2} & 2464 & 15$ ^{+} $ & 14$ ^{+}$   & 0.50(9)     & -0.19(6)  & $M1$\\
490.8\tnote{3} & 217.8(79) & 225.4(82)   & 1092 & 11$ ^{+}$  & 9$ ^{+}$    & 1.19(2)\tnote{1}  & +0.38(15)  & $E2$\\
537.9\tnote{3} & 53.6(33)  & 59.7(37)     & 1630 & 12$ ^{+}$  & 11$ ^{+}$   & 0.43(7)                  & -0.28(10)  & $M1$\\
588.3\tnote{3} & 33.0(26) & 36.0(29)     & 1680 & 12$ ^{+}$  & 11$ ^{+}$   & 0.55(10)                 & -0.17(3)  & $M1$\\
600.7\tnote{3} & 978(40)\tnote{4} & 1000(41) & 601  & 9$ ^{+}$   & 7$ ^{+}$ & 1.13(8)             & +0.40(5)  & $E2$\\
601.2                 &                       &    & (1202) & (11)   & (9) &  &  & \\
634.7 & 20.3(13)  & 20.4(13)      & 2445 & 14$ ^{-}$  & 13$ ^{+}$   & 0.55(9)                  & +0.18(9)  & $E1$\\
693.6 & 18.3(20) & 19.4(22)     & 1896 & 12$ ^{-}$  & 11$ ^{-}$   & 0.51(9)                  & -0.55(14)  & $M1$\\
703.5 & 18.3(15) & 19.4(16)     & 2600 & 13$ ^{-}$  & 12$ ^{-}$   & 0.49(25)                 & -0.26(13)  & $M1$\\
717.8\tnote{3} & 47.4(32) & 48.1(33)     & 1810 & 13$ ^{+}$  & 11$ ^{+}$   & 1.11(23)                 & +0.17(9)  & $E2$\\
743.3 & 14.4(17) & 15.1(18)     & 3207 & 16$ ^{+}$  & 15$ ^{+}$   & 0.53(24)                 & -0.23(7)  & $M1$\\
785.7 & 12.30(85)  & 12.82(90)     & 3231 & 15$ ^{-}$  & 14$ ^{-}$   & 0.52(26)                 & -0.26(13)  & $M1$\\
\hline
131.3\tnote{3} & 245(29) & 1618(194)  & 4149 & 17$ ^{+}$ & 16$ ^{+}$    & 1.03(7)              &       &$M1$ & $-  $\\
177.4\tnote{3} & 76.3(77) & 259(27)    & 5451 & 22$ ^{+}$ & 21$ ^{+}$    & 1.15(23)            &         &$M1$ & 54(27) \\
196.3\tnote{3} & 74.6(76)  & 208(21)    & 5647 & 23$ ^{+}$ & 22$ ^{+}$    & 1.16(12)\tnote{5}   &   &$M1$ & 14(4)\\
207.3 & 43.4(61)    &  & 6070 & (24)      & (22) & 1.84(55)                   &         &($E2$) \\
220.4 & 36.1(59)      &  & 6582 & (24)      & (22)           & 1.75(35)                  &           &($E2$) \\
245.9\tnote{3} & 370(37) & 722(73)   & 5274 & 21$ ^{+}$ & 20$ ^{+}$    & 1.10(15)              & -0.24(4)  &$M1$ & 35(8) \\
285.1\tnote{3} & 612(26) & 1000(43)  & 4434 & 18$ ^{+}$ & 17$ ^{+}$    & 1.08(3)                & -0.20(7)  &$M1$ & 34(6)\\
295.1\tnote{3} & 621(63) & 979(100)  & 5028 & 20$ ^{+}$ & 19$ ^{+}$    & 1.07(11)               & -0.25(5)  &$M1$ & 15(4)\\
298.7\tnote{3} & 647(51)\tnote{6} & 1006(80)\tnote{6}   & 4733 & 19$ ^{+}$ & 18$ ^{+}$    & 1.14(3)\tnote{7}  & -0.25(5)  &$M1$ & 20(4)\\
308.5 & 7.5(20)    &  & 6965 & (24)      & (23)       & 1.05(4)                  &           &($M1+E2$) \\
%347.7\tnote{8} & 127(12) \\
362.7 & 367(35) & 487(48)  & 5096 & 20$ ^{+}$ & 19$ ^{+}$    & 1.05(7)                    & -0.11(3)  &$M1$ \\
373.4 & 39.6(40) & 85(7)  & 5647 & 23$ ^{+}$ & 21$ ^{+}$      & 2.09(54)                   & +0.11(4)  &$E2$ \\
415.6 & 74(11) & 78(11)\tnote{9}  & 4434 & $18^+$ & $16^+$  & 1.85(56)                       & +0.68(28)  & $E2$ \\
422.5 & 19.7(27) & 20.7(29)  & 5451 & 22$ ^{+}$ & 20$ ^{+}$      & 1.82(81)\tnote{5}     & +0.59(19) &$E2$ \\
442.2 & 216(21) & 258(25)   & 3042 & 14$ ^{-}$ & 13$ ^{-}$    & 1.10(11)                  & -0.16(8)  &$M1$ \\
447.2 & 56.1(86)  &       & 5475 & 21        & 20$ ^{+}$   & 1.14(12)                  &           &($M1+E2$) \\
454.5 & 101(15) &     & 6838 & (24,25)       & (22,23)        & 1.74(20)                  &          &($E2$) \\
476.4\tnote{8} & 56(11)    &  &  &           &                   &           & \\
%496.8\tnote{8} & 120(17)      &    &         &                        &           & \\
499.9 &  150(19)\tnote{10}  &  & 7997 & (26,27)      & (25,26) & 1.36(21)         &           & ($M1+E2$) \\
541.2 & 43.1(77) & 44.3(79)    & 5274 & 21$ ^{+}$ & 19$ ^{+}$    & 1.96(44)\tnote{7} & +0.35(5)  & $E2$ \\
583.8 & 217(35) & 223(36)\tnote{9}  & 4733 & 19$ ^{+}$ & 17$ ^{+}$    & 1.98(36)\tnote{5} & +0.21(7)  & $E2$ \\
588.9 & 90.8(91)  &  & 5863 & (22)        & 21$ ^{+}$    & 1.01(43)                 &           & ($M1+E2$) \\
594.4 & 83.0(83) & 84.8(86)   & 5028 & 20$ ^{+}$ & 18$ ^{+}$    & 1.89(15)\tnote{7}& +0.34(6)  & $E2$ \\
617.8 & 128(20)   &  & 6383 & (22,23)      & (21)           & 1.56(48)                 &           &($M1/E2$) \\
%627.9\tnote{8} & 182(18) \\
658.8 & 132(19) &   & 7497 & (25,26)      & (24,25)         & 1.15(10)\tnote{7} &           &($M1)$ \\
668.9 &52.0(89)    &  & 5765 & (21)        & 20$ ^{+}$    & 1.20(31)                 &           & ($M1+E2$) \\
%678.8\tnote{8} & 124(16) & %144(19) & 4018 & 16$ ^{+}$ & 14$ ^{-}$    & 1.99(67)                & -0.23(7)  & $M2$ 
%\\
700.1 & 160(18)  & 169(19)    & 3300 & 14$ ^{-}$ & 13$ ^{-}$    & 1.05(29)                 & -0.37(10)  & $M1$ \\
718.3 & 205(23) & 233(26)   & 4018 & 16$ ^{+}$ & 14$ ^{-}$    & 2.00(93)                 & -0.31(3)  & $M2$ \\
737.2 & 29.8(45) &  & 5765 & (21) & 20$ ^{+} $ & \\
%739.4\tnote{8} & 86(11) &  &  & %14$ ^{-}$ & 13$ ^{-}$    & 1.09(50)                 & -0.27(15)  & $M1$  
%\\
774.6 & 62.1(68)   &  & 6638 & (23,24)      & (22)           & 1.36(49)                 &           & ($M1+E2$) \\
793.4 & 25.7(67)   & & 6656 & (23)      & (22)     & 1.23(34)                 &           & ($M1+E2$) \\
886.5 & 55(10)   &  & 6362 & (22)        & (21)           & 1.12(18)                 &           & ($M1+E2$) \\
975.6 & 185(22) & 196(24)    & 4018 & 16$ ^{+}$ & 14$ ^{-}$    & 2.00(58)                & -0.49(17)  & $M2$ \\
\hline 
\end{tabular}
%\end{center}
%\tablenote{Uncertainties in $E_\gamma$ are within $\pm (0.8 - 1.4)$ keV.}
\begin{tablenotes}
\item[1] From stretched quadrupole 601-keV ($E2$) gate.
\item[2] Transition was probably part of an inseparable doublet with another unplaced transition in coincidence with 601-keV ground state transition.
\item[3] Transitions reported in Ref.~\cite{hart}.
\item[4] Intensity for the composite peak of 600.7 and 601.2 keV.
\item[5] From pure dipole 295-keV ($M1$) gate.
\item[6] Transition in raw projection is inseparable from contaminant 298-keV transition of $ ^{205}$At.
\item[7] From pure dipole 285-keV ($M1$) gate.
\item[8] Transition could not be placed in the level scheme.
\item[9] Intensity normalized to that of the parallel $M1$ transition.
\item[10] Overlap with nearby strong 497-keV transition.
\end{tablenotes}
\end{threeparttable}
\label{Tb2}
\end{table*}

\begin{figure}[htbp]
%\vskip 6.0cm
%\hskip -4.0cm
%\begin{center}
\includegraphics[scale=.35,angle=0]{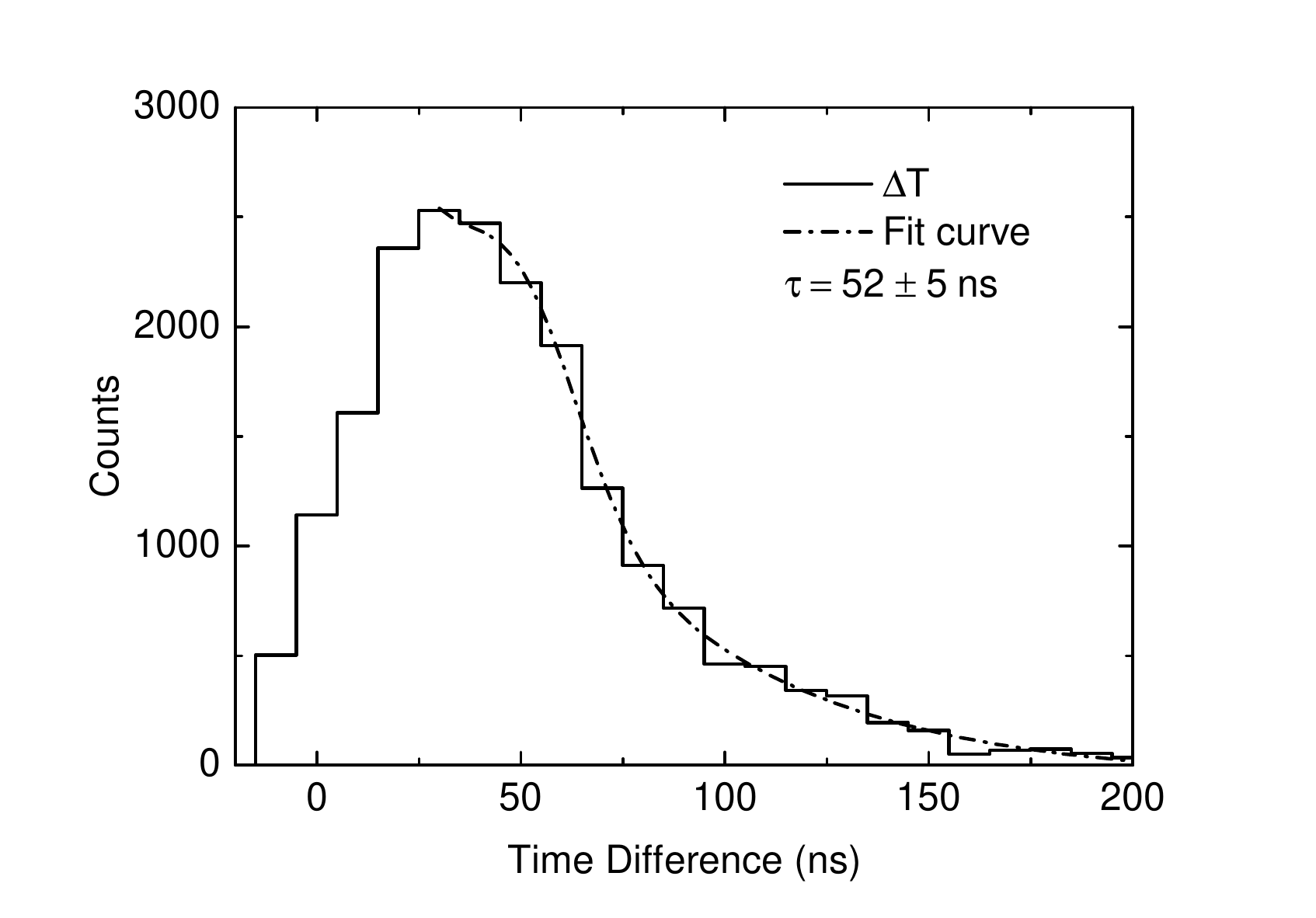} % Fig8
%\end{center}
%\vskip -08.2cm
\caption{ Time difference spectrum constructed using 285-, 299- and 295-keV transitions 
as {\em start} and 442- or 700-keV as {\em stop}. The dash-dotted curve is the exponential plus a Gaussian function fit to the peak and the decay tail of the spectra. The lifetime of the isomeric state is determined from the fit.}
\label{isomer}
%\end{center}
\end{figure}
%\vskip .5cm

Search for an isomeric state, as found from the missing intensity balance, was carried out by generating time-difference spectra, constructed by taking the 285-, 299-, and 295-keV transitions above the $ 16^{+} $ level as start and the 442- or the 700-keV transitions below the $ 16^{+} $ as the stop signal. The spectra showed delayed nature in the form of asymmetric tail of a Gaussian distribution. As statistics in the individual spectrum was limited, a few of these spectra were added and shown plotted in the Fig.~\ref{isomer}. A lifetime of $52(5)$~ns was extracted through exponential fit on the decay tail of the gated $ \Delta T $ spectra for the 4018-keV level (Fig.\ref{isomer}). This method is not suitable for finding isomers with lifetimes $\lesssim 30$~ns because of limitations in the dwell time and the intrinsic time resolution of the HPGe Clover detectors. Similarly, limitations on the coincidence time window had limited our search for long lived-isomers with $\tau \gtrsim 200\,{\rm ns}$.  

\section{Interpretation}
\label{intpret}
We take the clue from the fact that the doubly-odd At nuclei near the shell or sub-shell closure $(Z \approx 82, N \approx 120)$ provide a rich field of interplay between a few particle multiconfiguration shell model states and an even-even core with excitation. For interpretation purpose, the core acts as nearly magic nucleus. It has been demonstrated for different nuclei near the doubly magic configuration that the excitations are indicative of their single particle nature with separations providing strong evidence of the existing shell gap. Even for nuclei in the vicinity of $Z=82$ and $N=126$, shell model estimations by considering an inert core excitations involving a few particles (both protons and neutrons) and corresponding holes across the shell gap, have been attempted\cite{Fan2}. However, shell model calculations for $^{204}$At ($Z=85, N=119$) nucleus, with 3 proton particles and 7 neutron holes, become quite complicated. Furthermore, onset of collective behaviour, specially the formation of magnetic dipole band is found or expected to occur in similar nuclei (eg.\ $^{202}$Bi\cite{clar3}). However, possible configuration of the low lying yrast levels as well as the high-spin non-yrast states can be assigned based on our experimental data and related analysis, systematics of neighbouring nuclei and earlier results\cite{hart}.

As mentioned before, the $7^+$ ($\pi(1h_{9/2}) \otimes \nu(2f_{5/2}^{-1})$) ground state  and the $10^-$ ($\pi(1h_{9/2}) \otimes \nu(1i_{13/2}^{-1})$) first excited isomeric state of $^{204}$At were reported earlier\cite{gip}. The main sequence of 601-491-718-keV $\gamma$-ray transitions, reported earlier\cite{hart} was also confirmed in our measurements and firm spin-parity assignments were done based on our DCO and PDCO measurements(see Fig.~\ref{level}). In addition, a significant number of linking new transitions were established and placed in the level scheme. Most importantly, one isomeric transition and some of the crucial $E2$ crossover transitions were established through  our experiment to confirm the shears band structure at high-spin in $^{204}$At. Interpretation of these observations through the geometric model was also done, which will be discussed later (see Sec.~\ref{shears}). In the following sections, interpretation of different types of transitions and their sequences are discussed along with suggested predominantly single particle configurations and weak collectivity of $^{204}$At as manifested from our analysis and results from the neighbouring nuclei.

\subsection{Low lying states below {\protect $16^+$}}
\label{lowlying}
Configuration of the $7^+$ ground state with one proton in $1h_{9/2}$ suggests that two more $1h_{9/2}$ protons may be brought to the valence shell by breaking pairs leading to the configuration: $(\pi(1h_{9/2})^3 \otimes \nu(2f_{5/2}^{-1})$, which can account for the level sequence in the range $9^+$ to $13^+$ (marked as (I) in Fig.~\ref{level}). The negative parity states feeding the $13^+$ level may have a possible configuration: $(\pi(1h_{9/2})^3 \otimes \nu(1i_{13/2}^{-1})$ leading to maximum spin $I_{\rm max}=17^-$ and marked as (II) in Fig.~\ref{level}. This would involve promoting a neutron hole from $f_{5/2}$ to $i_{13/2}$, without changing the relatively robust proton configuration.

However, two more sequence of transitions were found to feed the $11^+$ state. These are marked as (III) and (IV) in the level scheme in Fig.~\ref{level}, and are likely to belong to different single particle configurations. One of the sequences leads to the $\Delta I=1$ band lying above the the $16^+$ level.
For this sequence marked by (III), a tentative assignment of configuration: 
$(\pi(1h_{9/2})^2\,\pi(1i_{13/2}) \otimes \nu(2f_{5/2}^{-1}))$ would account for the level sequence till $I_{\rm max}=17^-$. A change of configuration is expected to happen above in the same sequence leading to the possible configuration: $(\pi(1h_{9/2})^2 \pi(1i_{13/2}) \otimes \nu(1i_{13/2}^{-1})$, populating levels up to $21^+$. Existence of isomeric transition ($\tau= 52 \pm 5 \,{\rm ns}$) corresponding to $16^+ \rightarrow 14^-$ is expected to be of magnetic quadrupole type, as inferred from our isomer lifetime,  DCO and PDCO measurements (see Sec~\ref{expt}). Weisskopf single particle (SP) estimate of the lifetime based on the $M2$ transition is $\sim 1 - 10 \,{\rm ns}$, which is one order of magnitude lower than the experimental result, but closer to the SP value than for the other probable multipolarities. The deviation of the measured lifetime from the SP value may be attributed to the onset of collective behaviour in such nuclei with multiparticle configuration beyond the magic core. Incidentally, magnetic quadrupole $(M2)$ type transitions are found to occur in several neighbouring neutron deficient Astatine isotopes. Ground state transitions $({13/2}^+ \rightarrow {9/2}^-)$ in both $^{201}$At\cite{201at} and $^{203}$At\cite{203at} are of $M2$ nature. Similarly, $E3$ type ground state transition $(10^- \rightarrow 7^+$) was already observed in $^{204}$At\cite{gip}, which is another common feature in neutron deficient Astatine nuclei.     

The other level sequence marked as (IV) may be built up from the configuration: $(^{202}{\rm Pb} (4^+) \otimes \pi(1h_{9/2})^3 \otimes \nu(2f_{5/2}^{-1}))$, leading to 
$I_{\rm max}=17^+$.
Based on our observation of coincident K $X$-ray intensities and theoretical estimates, the 192-keV and the 167-keV $M1$ transitions are expected to be significantly converted. 
Our search for isomeric nature of the associated levels did not yield any conclusive result. Though the 588.3-keV $M1$ transition feeding the 1092-keV level could be observed in our experiment, no further transition feeding the 1680-keV $12^+$ level could be found. No attempt was made to assign configuration to the feeding level. 

\subsection{States above $16^+$}
\label{highlying}
As stated in the previous section, the isomeric $16^+$ level has been interpreted to originate from the configuration involving coupling of $i_{13/2}$ proton to aligned $h_{9/2}$ proton pairs, along with neutron hole in $i_{13/2}$, leading to: $(\pi(1h_{9/2})^2\,\pi(1i_{13/2}) \otimes \nu(1i_{13/2}^{-1}))$ configuration. It is worth mentioning that the same proton configuration was found to be responsible for the isomeric ${29/2}^+$ state in $^{203}$At \cite{203at}. Taking clue from this observation, it may be suggested that the above mentioned high-$j$ protons and the $i_{13/2}$ neutron hole will form the two angular momenta components contributing to the shears band observed in our experiment, with $16^+$ as band head.

Quite a few $\gamma$-ray transitions are observed around the shears band. These are placed in the level scheme (see Fig.~\ref{level}) which is extended up to $ \sim 8\, {\rm MeV}$ of excitation energy. Based on the feeding levels and our DCO and PDCO measurements, the spin parity assignments are partly done for these levels. Some of these transitions are found to be feeding the levels belonging to the shears band. No attempt is made to assign configuration to the high-spin states due to incomplete information available.    

\subsection{Shears band in $^{204}$At}
\label{shears}
\begin{figure}[htbp]
%\vskip 6.0cm
\begin{center}
%\hskip -0.5cm
\includegraphics[scale=.50,angle=0]{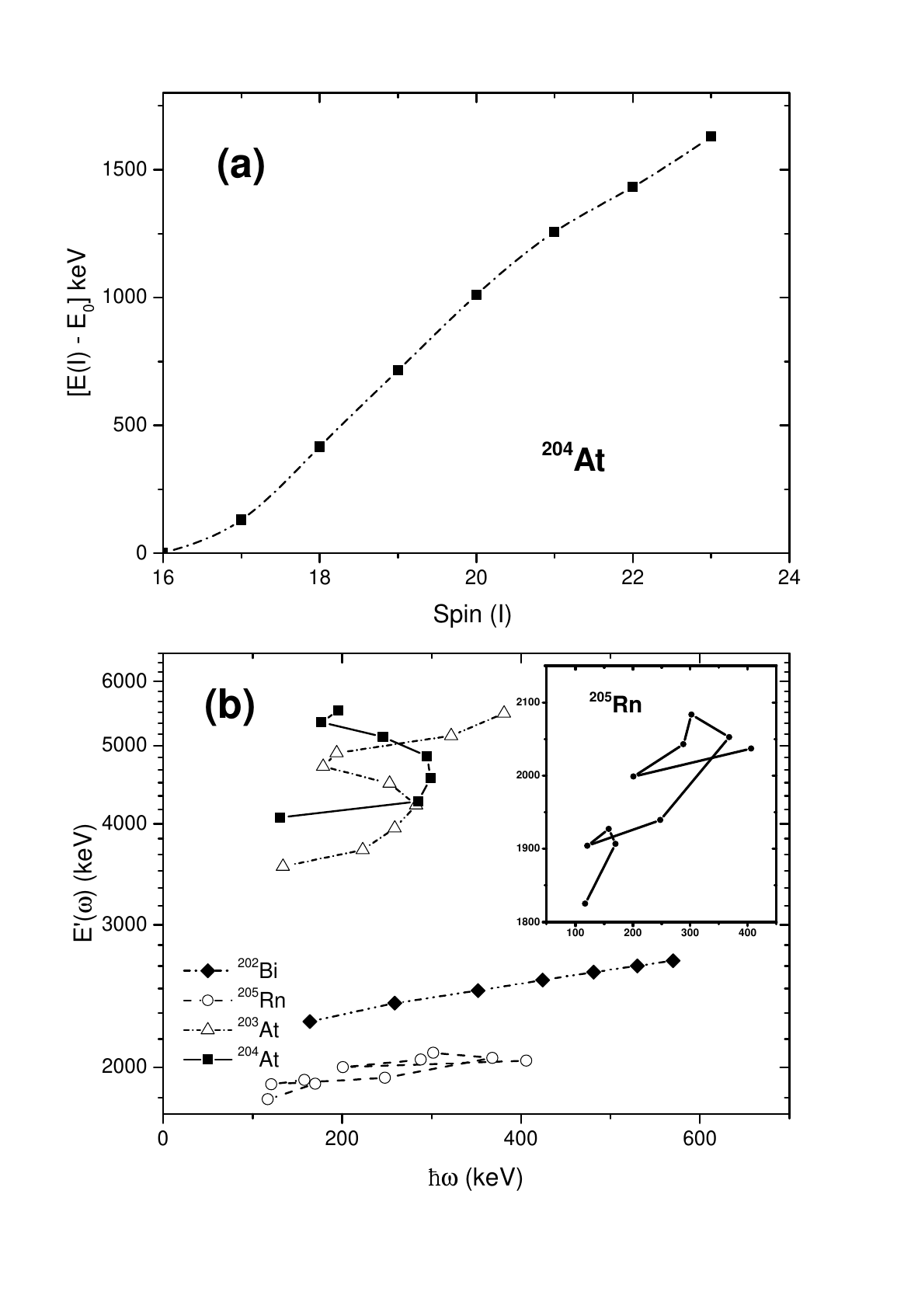} %Fig9
\end{center}
\vskip -1.2cm
\caption{(a) Energy of the levels of shears band with reference to the band-head is plotted as function of spin $(I)$. The dash-dotted line through the points is drawn to guide the eye. It also shows a kink around $I=21$ indicating band crossing. (b) Plots of experimental routhians for $^{204}$At, $^{203}$At\cite{203at}, $^{205}$Rn\cite{nova} and 
$^{202}$Bi\cite{clar3}. Scale shift along $E'(\omega)$ axis would occur due to uncertainties in energy of the associated levels due to unobserved transitions in $^{202}$Bi, $^{205}$Rn and $^{203}$At. The lines are drawn to guide the eye. The inset shows the $^{205}$Rn plot over expanded scales.}
\label{rou}
%\end{center}
\end{figure}

Energy levels of the shears band $(I^\pi=16^+ - 23^+)$, with band-head energy $(E_0)$ as the reference, are plotted as function of the spin in the Fig.~\ref{rou}~(a), which clearly shows the expected quadratic pattern of a band. In addition, the nucleus was found to be weakly oblate with the quadrupole deformation parameter ($\beta_2=-0.084$), as estimated from the finite range droplet model (FRDM)\cite{andt}. Another estimate based on PES calculations using relativistic mean field (RMF) theory with NL3 effective interaction\cite{lian} predicts somewhat larger value ($\beta_2=-0.143$), though it indicates predominantly oblate deformation as well. Furthermore, based on earlier observations\cite{hart}, and our measurements of DCO and PDCO ratios of the transitions belonging to the sequences above $16^+$, magnetic dipole nature of the band is confirmed. Weak $E2$ cross-over transitions were clearly evident as interspersed within the band. In addition, high $j$ orbitals are involved as the active orbitals, resulting in weak oblate deformation and the shears band structure involving two blades of a shear, possibly along with coupling of the core angular momentum. 

Existence of dipole bands are found in many of the neighbouring isotones of $^{204}$At ($N=119$). First observation of low energy magnetic dipole transitions between levels at high-spin were first observed in the isotope $^{202}$Bi\cite{clar3}, followed by observation in $^{205}$Rn\cite{nova}. While the spacings of levels are regular in $^{202}$Bi, 
those of $^{205}$Rn were found to be irregular. These are shown in the experimental routhian (ER) plots of Fig.~\ref{rou}. Recently, observation of similar $M1$ band with somewhat irregular spacings was reported in $^{203}$At\cite{203at}. ER plot of $^{203}$At is also shown in the Fig.~\ref{rou}, which indicates similarity with that of $^{204}$At. 

%\vskip .5cm

The MR bands identified in $^{202}$Bi were suggested as arising from $2p-1h$ proton configuration, with at least one of the protons in high-$j$ $h_{9/2}$ or $i_{13/2}$ orbital, coupled to one or more high-$j$ $ i_{13/2}$ neutron holes. Estimated band head spins due to the coupling is expected to cover the range: $I = 10 - 16 \, \hbar$ of the band indicating that one configuration can build up the dipole band. No evidence of back-bending or band-crossing resulting from change of particle hole configuration is seen in $^{202}$Bi.

On the other hand, evidence of band crossing was reported in $^{205}$Rn\cite{nova}, which is included in the Fig.~\ref{rou}. Based on TRS calculations, two likely candidates for configuration of the dipole band of positive and negative parity were suggested. Band crossing due to alignment of $i_{13/2}$ neutron hole pairs around $\hbar \omega\sim 0.3\, {\rm MeV}$ was evident. Though it is not supported by the large $B(M1)/B(E2)$ ratios for the lower rotational frequencies ($\hbar \omega < 0.2 \, {\rm MeV}$) of the dipole band, another band crossing at $\hbar \omega\sim 0.15 \, {\rm MeV}$ might have occurred indicating another change of configuration or alignment of particle-hole pairs.

Back bending is clearly evident in Fig.~\ref{rou} along the dipole bands for both $^{203}$At and $^{204}$At at $\hbar \omega\sim 0.25 - 0.3 \, {\rm MeV}$. In $^{204}$At, this can be attributed to band-crossing resulting from change of particle-hole configuration. As mentioned in Sec.~\ref{highlying}, the lower spin part of the band is accounted for by the  $\pi(1h_{9/2})^2\,\pi(1i_{13/2}) \otimes \nu(1i_{13/2}^{-1})$ configuration, while post band crossing configuration would arise from breaking of two more particle-hole pairs leading to 
$\pi(1h_{9/2})^4\,\pi(1i_{13/2}) \otimes \nu(1i_{13/2}^{-1})$. Core rotation is expected to play an important role here. Agreement or disagreement with the above mechanism at play can be made through theoretical calculations of observables such as: $B(M1)$ and $B(M1)/B(E2)$ for the shears band using the semiclassical or geometric model\cite{clar2} called shears mechanism with principal axis cranking (SPAC) model and compare with the experimentally extracted quantities. Details of the SPAC calculations and corresponding experimental results are discussed in the following section.

\subsection{SPAC calculations for dipole band and interpretation of results}
\label{spac}
Details of the formalism of SPAC model with core rotation is given in Ref.~\cite{spac}. According to the energy systematics of the model, the total energy $E(I)$ for a given angular momentum $I$: $(\vec{I}={\vec{j}}_{sh}+\vec{R})$, where ${\vec{j}}_{sh}$ is the shears angular momentum and $\vec{R}$ is the core angular momentum contributing to the shears mechanism, is expressed as: \\
\begin{equation}
\nonumber
E(I)=E({\rm shears}) + E({\rm core}) + E_0,
\end{equation}
where $E_0$ accounts for the potential energy, $E({\rm shears})$ and $E({\rm core})$ are the excitation energies imparted to the system through shears mechanism involving quasiparticle interaction and collective core rotation respectively. In case of nuclei around Lead region ($Z \sim 82$), the two blades of the shears can be separated into proton particles constituting one blade and neutron holes ($N < 126$) forming the other blade.  The shears mechanism involves orientation of the two blades of shear carrying quasiparticle angular momenta $\vec{j_\pi}$ and $\vec{j_\nu}$, such that $\vec{j_{sh}}= \vec{j_\pi}+\vec{j_\nu}$. Considering the rotation axis along $\hat{x}$, the core rotation angular momentum vector $\vec{R}$ must be oriented along x-axis. If $\theta_1$, $\theta_2$ are the tilt angle of $\vec{j_\pi}$, $\vec{j_\nu}$ respectively with the rotation axis $\hat{x}$, the magnitude of the core angular momentum can be written as: \\
\begin{align*}
\nonumber
R(I,\theta_1,\theta_2)=&\sqrt{I^2-(j_\pi\sin{\theta_1}+j_\nu\sin{\theta_2})^2} \\& - j_\pi\cos{\theta_1}-j_\nu\cos{\theta_2}.
\end{align*}
The core angular momentum $\vec{R}$ is expected to make a small but non-zero correction to the component of the contribution from the shears mechanism. 

Quasiparticle contribution $E({\rm shears})$ to shears mechanism is accounted for through spin independent effective interaction: $V(\theta)$, where $\theta=\theta_1-\theta_2$ is the angle between the shears. From symmetry consideration it is expressed as:\\
\beq
\nonumber
V(\theta)=V_0+V_2P_2(\cos{\theta}).
\eeq  
The above equation is the multipole expansion in terms of the shears angle $\theta$, where only the zeroth order (independent of $\theta$) and the second order terms are included. Based on the systematics of the effective interaction in $^{198,199}$Pb\cite{mac1}, predominance of the $P_2$ term in the $\theta$ dependence of the interaction potential is established. Furthermore, attractive nature of the particle - hole interaction as manifested in  $^{198,199}$Pb, require $V_2$ to be positive. Experimentally, $V_2$ value acceptable for the Pb region is $\sim 2.3 \,{\rm MeV}$ with approximately 2 - 4 proton and neutron holes participating in the shears band formation, resulting in $\sim 300 \,{\rm keV}$ contribution to $V_2$ per proton / neutron hole pair.

\begin{figure}
\vskip -1cm
%\hskip -4.0cm
\begin{center}
\includegraphics[scale=.35,angle=0]{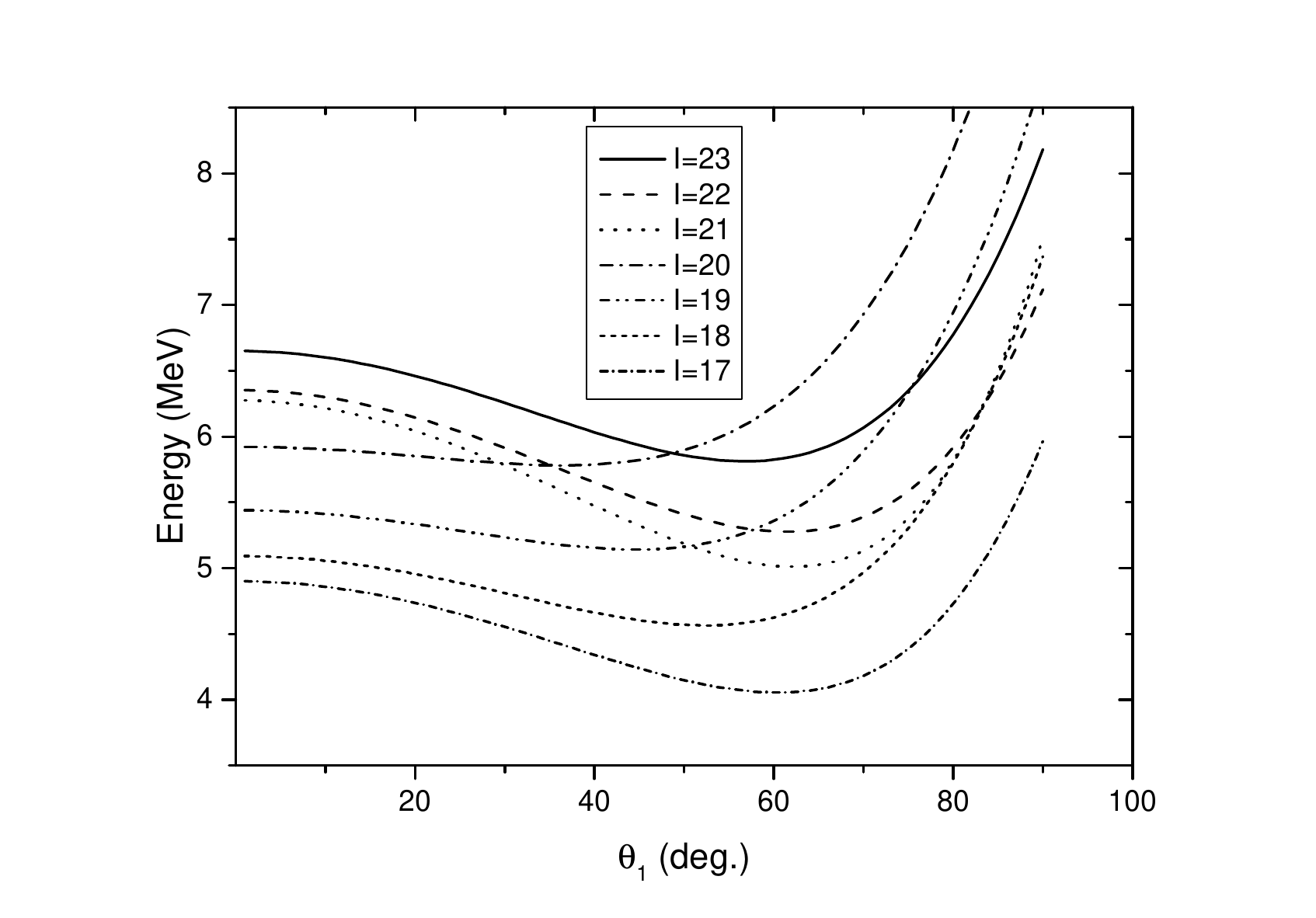} % Fig10
\end{center}
\vskip -1cm
\caption{Energy minimization plots according to SPAC model based estimates for the MR band of $^{204}$At.}
\label{minmiz}
%\end{center}
\end{figure}
%\vskip .5cm
On the basis of the SPAC model, energy minimization for each $I$ is next achieved to extract $\theta_1$ and $\theta_2$\cite{clar2}. Since the ${1i_{13/2}}^{-1}$ configuration for the neutron hole blade is considered throughout the band, we assume $\theta_2=0$ as fixed for the shears band. In other words, only the proton blade alignment is considered. By this assumption, the energy minimization condition boils down to one dimension: \\
\beq
\nonumber
{\left(\frac{\partial E}{\partial \theta_1}\right)}_I=0.
\eeq
The level energy $E(I)$ at minimization condition is extracted. Energy minimization plots ($E(I, \theta_1) \, {\rm vs.} \, \theta_1$) for different levels belonging to the dipole band are shown in the Fig.~\ref{minmiz}. Band-crossing above $I = 20 \, \hbar$ can be clearly seen.

The experimental observables such as the magnetic dipole transition probabilities $B(M1)$ and crossover electric quadrupole transition probabilities $B(E2)$ along the band are calculated as:\\
\begin{align*}
B(M1, I \rightarrow I-1)  = {\frac{3}{8\pi}}& [{g_\pi}^* j_\pi \sin(\theta_1-\theta_I)\\& - {g_\nu}^* j_\nu \sin(\theta_I-\theta_2)]^2, 
\end{align*}
\beq
\nonumber
B(E2, I \rightarrow I-2) = {\frac{15}{128 \pi}} \left[Q_{\rm eff}\sin^2 \theta_{\pi j}+ 
Q_{\rm coll} \cos^2 \theta_I \right]^2, \\
\eeq
where ${g_\pi}^*=g_\pi - g_R, \, {g_\nu}^*=g_\nu - g_R, \, g_R=Z/A\,$
and $\theta_I$ is the angle subtended by $\vec{I}$ with the rotation axis $\hat{x}$, $\theta_{\pi j}$ is the angle subtended by $\vec{j_\pi}$ with $\vec{j_{\rm sh}}$, $g_\pi$, $g_\nu$ are the effective g-factors of the proton and the neutron hole blades respectively, and $Q_{\rm eff}$, $Q_{\rm coll}$ are the quasiparticle and collective quadrupole moments respectively. Values of these crucial parameters are extracted from the available data for the nuclei around $Z=82$. These are given in the following sub-section~\ref{para}.    

\subsection{Quasiparticle configurations and choice of parameters}
\label{para}

As suggested in Sec.~\ref{shears}, part of the band before band-crossing may be accounted for by the  $\pi(1h_{9/2})^2\,\pi(1i_{13/2}) \otimes \nu(1i_{13/2}^{-1})$ configuration, while post band-crossing configuration would arise from breaking of two more particle-hole pairs leading to $\pi(1h_{9/2})^4\,\pi(1i_{13/2}) \otimes \nu(1i_{13/2}^{-1})$. $j_\pi = 12.5 \, \hbar$ for the proton quasiparticle blade of unstretched configuration and $j_\nu=5.5 \, \hbar$ for the neutron hole are considered. Post band-crossing, proton configuration changes but unstretched value of $j_\pi = 16.5 \, \hbar$ is used, keeping the $j_\nu$ unchanged.  

\begin{figure}
\vskip-1.0cm
%\hskip -1.0cm
\begin{center}
\includegraphics[scale=.50,angle=0]{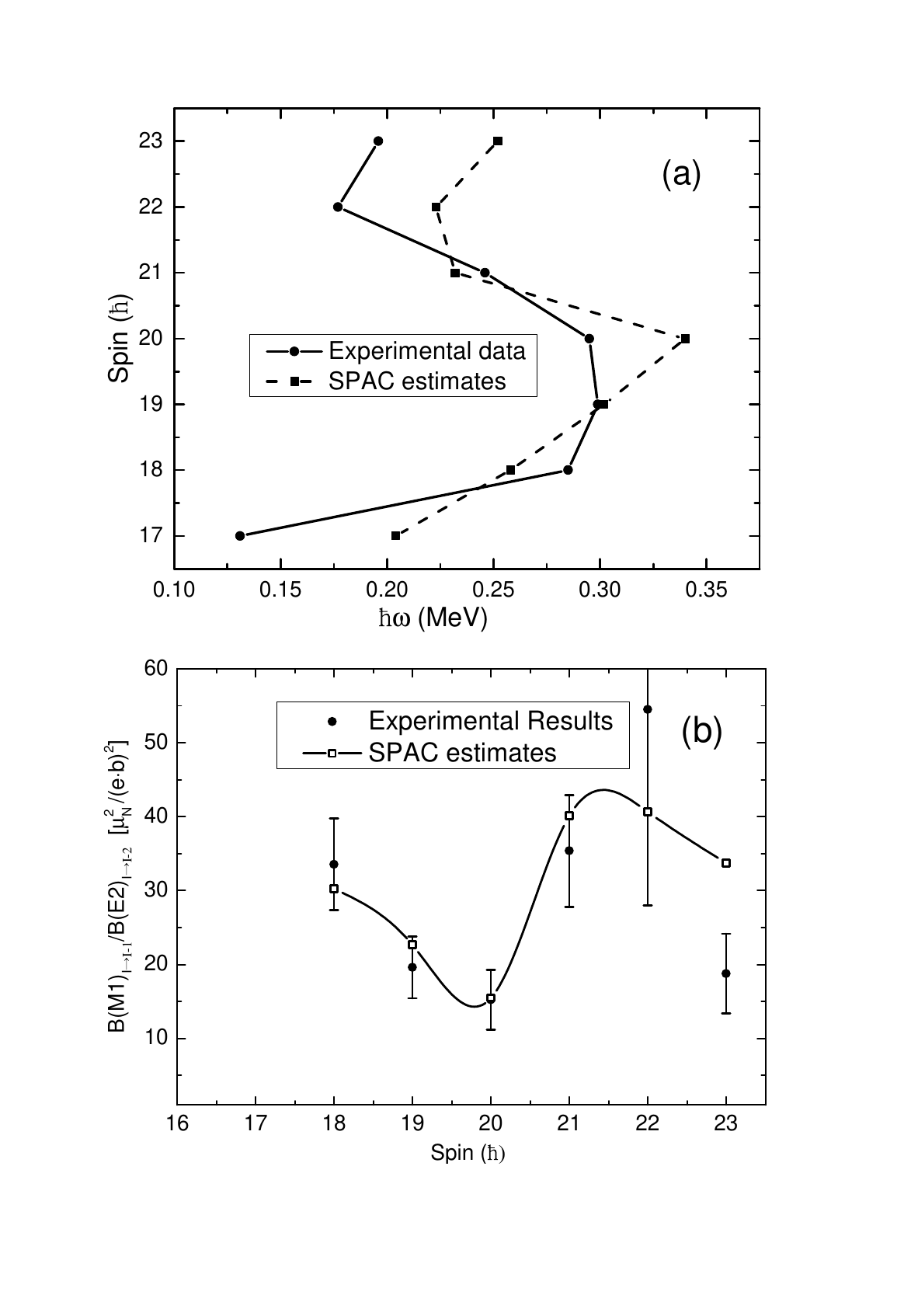} % Fig11
\end{center}
\vskip -2.2cm
\caption{(a) Plot of level spin versus rotational energy of the levels belonging to the shears band in $^{204}$At. SPAC model based estimates are also plotted. The line joining the data points are also shown. (b) The $B(M1)/B(E2)$ values for the shears band, extracted from the experiment, are compared with the SPAC estimates based on the chosen or adjusted parameter values (see text). The continuous line is spline interpolated through the SPAC estimates to guide the eye.}
\label{m1e2}
%\end{center}
\end{figure}
%\vskip .5cm
Weak oblate deformation is observed in Astatine isotopes (eg. $^{208}$At ($J^\pi=10^-$), with $Q \sim - 1.67\, eb$ \cite{andt1}. $Q_{\rm coll}=1.8 \, eb$  before band-crossing is used in our SPAC estimates to account for the collective core rotation. The effective quadrupole moment due to quasiparticle configuration is given by \cite{clar2}: $Q_{\rm eff} = \left(\frac{e_\pi}{e}\right)\,Q_\pi + \left(\frac{j_\pi}{j_\nu}\right)^2 \,\left(\frac{e_\nu}{e}\right)\, Q_\nu$, where $Q_\pi, \, Q_\nu$ are the quadrupole moments of the valence protons and neutrons bound within the nucleus and $e_\pi, \, e_\nu$ are the $E2$ polarization charges of the valence protons and neutrons. The polarization charge is given by $e_{\pi (\nu)}=(T_z+1/2)\, e + e_{\rm Pol}(E2)$, where $T_z$ is the third component of isospin with $T_z=\pm 1/2$ for protons and neutrons respectively, and $e_{\rm Pol}(E2)$ is the $E2$ polarization charge. For the Lead region, $e_{\rm Pol}(E2)\sim 3e$ is needed to match the experimental value\cite{clar2}. The extracted value of $Q_{\rm eff}$ from the experimental $B(E2)$ results for the dipole band in $^{198, 199}$Pb, is found to be $\sim 6.5 \, {\rm eb}$\cite{clar1}. Collective contribution due to core rotation is neglected in arriving at the above result and therefore, somewhat smaller value $Q_{\rm eff} \sim 4 \, {\rm eb}$ is chosen as trial value in our case for $^{204}$At to match with the experimental results. This value also agrees with the estimate of $Q_{\rm eff}$ based on the multiparticle configuration of the band.    

In the SPAC calculation, adjustment of the parameters, such as $V_2$  and the core moment of inertia $J(I)$ are done to reproduce the energy levels through minimization. Extracted parameter values are: $V_2=0.8 \, {\rm MeV}$ up to the band crossing spin $I=20 \, \hbar$. This is consistent with the experimentally extracted value of $\sim 0.3 \, {\rm MeV}$ per particle/hole pair contributing to the shears mechanism\cite{clar2}. Since the number of particle / hole pair increases above the band crossing, $V_2 = 1.8 \, {\rm MeV}$ is found to be reasonably consistent. Core moment of inertia $J(I)$ need to be increased at higher spin over the range $7.8 - 16.5 \, \hbar^2/{\rm MeV}$. Core contribution to the total angular momenta is found to be $\lesssim 20\%$, highest contribution being $\sim 17\%$ at 
$I=23 \, \hbar$. These results are in reasonable agreement with the systematics in $^{198,199}$Pb\cite{mac1}, where $5 - 15\%$ core contribution to the total spin was reported. The spin $(I)$, as function of the rotational energy for the shears band is plotted in the Fig.~\ref{m1e2}~(a). The results of SPAC minimization at each spin is adjusted to be in reasonable agreement with the experimental results. These are also plotted in the same figure.

The $g$-factors are used as input to calculate the magnetic dipole transition rates. These results are tabulated in Ref.~\cite{Hub}. The factors are obtained from time dependent perturbed angular distribution (TDPAD) experiments and found to be in good agreement with the shell model estimates. From the equivalent quasiparticle configuration, the adopted values used as input to SPAC estimates are: $g_\pi = 1.060(14)$ (from $^{209}$At) and $g_\nu = -0.150(6)$ (from $^{200}$Pb). 

The $B(M1)/B(E2)$ ratios, obtained from the SPAC model based estimates are compared with those results obtained from our experiment using the measured intensities of the $M1$ and corresponding cross-over $E2$ transitions (see Eq.~\ref{BM1BE2}). The mixing parameter $\delta$ is considered as negligible since the corresponding DCO ratios indicate negligible admixtures to the $M1$ transitions (see Fig.~\ref{dco2}). The results are shown in the Fig.~\ref{m1e2}~(b). The uncertainties in the experimental data are estimated from the intensity fitting errors, which are normally large because of weak cross-over $E2$ transitions. The SPAC model estimates along the shears band agree reasonably well, supporting the band crossing and the associated quasiparticle configurations constituting the shears band.  
 
\section{Summary}
\label{sum}
The excited states of doubly odd $^{204}$At, leading to high-spin states up to $\sim 8\, {\rm MeV}$ and $\sim 25\, \hbar$ have been observed from our experiment involving decay of the evaporation residue formed in $^{12}$C + $^{197}$Au fusion reaction with $^{12}$C as projectile at 65 and 75~MeV beam energies. The level scheme, which was partially known from earlier studies, was extended through results of DCO and PDCO measurements using the Clover Detector array INGA with 15 detectors. Quasiparticle configurations involving valence proton particles and neutron holes in respective high $j$ orbitals are found to be able to account for the level scheme. One isomeric transition of magnetic quadrupole type is found through our isomer search over a limited range of time delay available and time resolution due to time stamping of on-line data with 10~ns tick of the real time clock in our experimental arrangement. Isomers with lifetime: $30 \, {\rm ns} \lesssim \tau \gtrsim 200\, {\rm ns}$ would be difficult to explore with limited statistics. Mismatch of intensity was found in a few cases, but the isomer search remained inconclusive. The lifetime of the $M2$ transition was found to be in  agreement within the approximations for the single particle Weisskopf estimates. 

Existence of shears band structure at high-spin was observed earlier in Gammasphere experiment\cite{hart}, but it is confirmed in our experiment with observations of the cross over $E2$ transitions and the spin parity assignments based on the DCO and PDCO ratio measurements. The experimental $B(M1)/B(E2)$ ratios for the dipole band are shown to be in good agreement within the uncertainty limit with theoretical estimates based on the SPAC model, which takes care of the shears mechanism involving participating valence nucleons (particles and / or holes) at the high-$j$ orbitals. The shears band  structure is found to be due to participation of proton particles in the $1h_{9/2}$ and / or $1i_{13/2}$, and neutron hole in $1i_{13/2}$. Multiple protons are expected to contribute to the structure formation. Band crossing along the shears band have also been observed and accounted for on the basis of SPAC estimates. 

Although the magnetic dipole bands were first found in the neutron deficient Lead isotopes with involvement of multiple $1i_{13/2}$ neutron holes, a significant number of other nuclei around $Z=82$, such as $^{202}$Bi, $^{205}$Rn and $^{203,204}$At also manifest similar nature of the nuclear structure at high-spin.  Similarities and differences in the nature of these dipole bands were illustrated and the results were found to be in reasonable agreement with the model based on shears mechanism, as proposed by Machhiavelli et al.\cite{mac2}. 

\section{Acknowledgement}
Authors would like to express their gratitude to the staff of the BARC-TIFR Pelletron Accelerator facility for smooth running of the machine during the experiment. Special thanks go to the members of the INGA collaboration for setting up the Clover Detector Array. We are grateful to S. Rajbanshi and Sajad Ali for their vauable suggestions on the theoretical estimates. We acknowledge the financial support for the project by the Department of Atomic Energy, Government of India, under grant no. 12-R \& D-SIN-5.02-0102.

\end{document}